\documentclass[3p,number,sort&compress,fleqn,preprint]{elsarticle}
\usepackage{amssymb,amsmath,latexsym}
\usepackage{pifont}   
\usepackage{natbib}   
\usepackage{geometry} 
\usepackage{fleqn}    
\usepackage{graphicx} 
\usepackage{color}
\newcommand{\ii}{\mathop{\text{i}}}
\newcommand{\ee}{\mathrm{e}}
\newcommand{\dd}{\text{d}}
\renewcommand{\d}{\text{d}}

\newcommand{\Aplus}{A^{\text(+)}}
\newcommand{\Aminus}{A^{\text(-)}}
\newcommand{\Apm}{A^{\text(\pm)}}
\newcommand{\sfAplus}{\mathsf{A}^{\text(+)}}

\newcommand{\bbeta}{\boldsymbol{\beta}}
\newcommand{\bhbeta}{\boldsymbol{\widehat{\beta}}}

\newcommand{\betaT}{\beta_\text{T}}
\newcommand{\betaL}{\beta_\text{L}}

\newcommand{\cT}{c_\text{T}}
\newcommand{\cL}{c_\text{L}}

\newcommand{\bM}{\mathbf{M}}

\newcommand{\bq}{\mathbf{q}}
\newcommand{\br}{\mathbf{r}}
\newcommand{\bQ}{\mathbf{Q}}
\newcommand{\bR}{\mathbf{R}}
\newcommand{\rr}{\mathbf{r}}
\newcommand{\bV}{\mathbf{V}}

\newcommand{\Bs}{\mathbf{s}}

\newcommand{\hk}{\widehat{k}}
\newcommand{\hr}{\widehat{r}}
\newcommand{\hR}{\widehat{R}}

\newcommand{\hn}{\widehat{n}}
\newcommand{\bhe}{\mathbf{\widehat{e}}}
\newcommand{\bhk}{\mathbf{\widehat{k}}}
\newcommand{\bhn}{\mathbf{\widehat{n}}}

\newcommand{\bhr}{\mathbf{\widehat{r}}}
\newcommand{\bhR}{\mathbf{\widehat{R}}}
\newcommand{\bhV}{\mathbf{\widehat{V}}}

\newcommand{\Giso}{G^{\text{iso}}}

\newcommand{\pv}{\mathop{\text{p.v.}}}
\newcommand{\Pf}{\mathop{\text{Pf}}}
\newcommand{\sign}{\mathop{\text{sign}}}
\newcommand{\norm}[1]{|\!|{#1}|\!|}
\renewcommand{\Re}{\mathop{\text{Re}}}
\renewcommand{\Im}{\mathop{\text{Im}}}

\begin{document}
\journal{Wave Motion}
\begin{frontmatter}
\title{On the gradient of the Green tensor in two-dimensional elastodynamic problems, and related integrals: Distributional approach and regularization, with application to nonuniformly moving sources}
\author[cea]{Yves-Patrick Pellegrini\corref{cor1}}
\ead{yves-patrick.pellegrini@cea.fr}
\address[cea]{CEA, DAM, DIF, F-91297 Arpajon, France.}

\author[darm]{Markus Lazar}
\ead{lazar@fkp.tu-darmstadt.de}
\address[darm]{Heisenberg Research Group,
        Department of Physics,\\
        Darmstadt University of Technology,
        Hochschulstr. 6,
        D-64289 Darmstadt, Germany.}

\cortext[cor1]{Corresponding author}
\date{\today}
\begin{abstract}
The two-dimensional elastodynamic Green tensor is the primary building block of solutions of linear elasticity problems dealing with nonuniformly moving rectilinear line sources, such as dislocations. Elastodynamic solutions for these problems involve derivatives of this Green tensor, which stand as hypersingular kernels. These objects, well defined as distributions, prove cumbersome to handle in practice. This paper, restricted to isotropic media, examines some of their representations in the framework of distribution theory. A particularly convenient regularization of the Green tensor is introduced, that amounts to considering line sources of finite width. Technically, it is implemented by an analytic continuation of the Green tensor to complex times. It is applied to the computation of regularized forms of certain integrals of tensor character that involve the gradient of the Green tensor. These integrals are fundamental to the computation of the elastodynamic fields in the problem of nonuniformly moving dislocations. The obtained expressions indifferently cover cases of subsonic, transonic, or supersonic motion. We observe that for faster-than-wave motion, one of the two branches of the Mach cone(s) displayed by the Cartesian components of these tensor integrals is extinguished for some particular orientations of source velocity vector.
\end{abstract}

\end{frontmatter}
\section{Introduction}
The two-dimensional elastodynamic Green tensor~\cite{EASO56} of the Navier equation is the primary building block of solutions for linear elasticity problems involving nonuniformly moving rectilinear line sources, such as dislocations (e.g., \cite{LAZA11,LAZA13a}). Dislocations are the fundamental carriers of plastic deformation in crystalline materials \cite{KRON58,LARD74,HIRT82}. Mathematically, a dislocation stands as a discontinuity of the displacement field on its glide plane. This discontinuity stands as a boundary condition in traditional methods of solution of continuum mechanics \cite{MARK01,GURR13}, or acts more explicitly as a singular source of elastic field if the solution is tackled by means of Green functions \cite{MARK83}. In the latter approach, the elastodynamic solution for the strain or stress fields involves taking convolution integrals of derivatives of the Green tensor by the source functions \cite{MURA69,MURA87}.

Whereas the two-dimensional Green tensor itself is locally integrable, its derivatives are in general hypersingular kernels \cite{MART96}, namely, kernels that cannot simply be regularized by means of a Cauchy principal-value requirement. Still, they are fully legitimate objects within the theory of distributions, their apparent singularity being handled by introducing Hadamard's finite-part prescription \cite{SCHW50,GELF64,ZEMA65,BLAN00,ESTR02,KANW04}. From an operational standpoint, finite parts can ultimately be reduced by carrying out suitable integration by parts on convolution integrals \cite{GURR13}. Hypersingular kernels are commonly encountered in situations involving static \cite{KARL00}, as well as nonuniformly moving dislocations or cracks \cite{COCH94,PELL10}, notably in the context of so-called boundary-element integral approaches \cite{KOLL92,MART96,DANG05}, and because of their importance in practice, their handling is a recurrent issue in wave physics problems \cite{GURR13,DANG05,ARDA99}.

In the two-dimensional setting, the dislocation line source is transverse to the $(x,y)$ plane of motion, in which it reduces to a point in the idealized case of a Volterra dislocation.
In general, point sources lead to singular (infinite) fields at the source position and at wave fronts, which poses some problems in numerical implementations \cite{GURR13}. However, infinite fields are merely the hallmark of the breakdown of classical linear elasticity at the dislocation core. In reality, a  physical dislocation has a finite width, that can be measured in atomistic simulations or computed by means of specially devised nonlinear models of the cohesive-zone type \cite{PEIE40,MRYA98,PELL13}. Also, dislocations of a finite width naturally arise in the framework of gradient elasticity models (e.g., \cite{LM05,LM06,LAZA12,LAZA13b,L14}).

Being of finite width is a necessary condition for sources to undergo supersonic motion (or faster-than-light-speed motion  in classical electrodynamics \cite{ARDA99}). Indeed, faster-than-wave motions of point sources induce Mach or Cerenkov cones with unrealistically infinite field strength \cite{ARDA99}. For dislocations or cracks, faster-than-wave motion \cite{WEER67} has attracted wide attention during the last decades \cite{PELL13,GUMB99,ROSA99,NOSE07,LAZA09,CALL80,MARK08,WEER14}. Also, recent medical imaging techniques rely on shear-wave Mach cones induced by a fast moving ultrasonic spot at the surface of human skin \cite{BERC04a,BERC04b}. Thus, supersonic motion must be allowed for in any comprehensive theory of radiation fields generated by moving sources. We should add that, quite generally, the concept of a point source can hardly be avoided when no information about the physical nature of the singular source of field is available. Then, Hadamard's finite part regularization, or generalizations thereof, must be employed. We refer the interested reader to Ref.\ \cite{BLAN00} for a review of some recent progresses in this direction, motivated by the problem of relativistic motion of a point particle in general relativity.

However, in the specific context of dislocation theory, convoluting the point source by an appropriate shape function of finite width that represents the core provides a natural regularization of the relevant field integrals at the source location, and at the wave fronts (including Mach cones), and allows one to investigate subsonic as well as supersonic motion without the need to address these cases separately \cite{PELL11}. In many approaches to finite-size (so-called \emph{smeared-out}) dislocations~\cite{ESHE49}, core shape functions are often found or assumed of power-law decay in the space variable \cite{LOTH92,MRYA98,CAIA06}. On the other hand, an exponentially-decaying shape function with cut-off characteristic length is produced by the theory of gradient elasticity of the Helmholtz type (e.g., \cite{LAZA13b,L14}) where the convolution is naturally embedded within the Green function of the theory as a consequence of the constitutive relations employed.

This paper introduces an alternative power-law-type way of regularizing the elastodynamic dislocation problem, which tames all singularities of the fields in the whole $(x,y)$ plane. While resembling certain means \cite{CAIA06} currently employed to regularize elastostatic fields in three-dimensional simulations \cite{CAIA06,Po13} it will arise, however, from an immediate analytic continuation of the Green tensor to complex values of the time variable, once the elastodynamic fundamental solutions are written down as distributions. Simplicity of implementation is indeed a necessary requirement for use in dislocation-dynamics simulations \cite{GURR13,CAIA06}.

Section \ref{sec:greentens} reviews several different forms of the two-dimensional elastodynamic Green tensor of the Navier equation for the material displacement in an isotropic medium, and its derivatives, which we express as distributions. Their regularization is examined in Sec.\ \ref{sec:reggf}, and applied in Sec.\ \ref{sec:movingline} to the computation of specific key definite integrals over time, that enter the problem of sources undergoing a velocity jump from rest to an arbitrary constant velocity, in the plane-strain and anti-plane-strain settings relevant to screw and edge dislocations, respectively.
These key integrals ---from which expressions of the strain and stress fields can be deduced \cite{LAZA14}--- lead, when employed for faster-than-wave source motion, to Mach cones which we further analyze here in terms of distributions. The key integrals are obtained as a difference at their time boundaries of non-trivial indefinite integrals. The latter can be used to address more general nonuniform source motions since for numerical purposes, a nonuniform motion can in general be represented conveniently as a succession of velocity jumps separating small time intervals of uniform motion \cite{PILL07,PELL13}. The full solution for the fields in this problem will be reported elsewhere \cite{LAZA14}.
A discussion (Sec.\ \ref{sec:concl}) closes the paper.

\emph{Notations:} Throughout the text, the `hat' notation is employed to denote the unit director $\mathbf{\widehat{a}}=\mathbf{a}/a$ of a vector $\mathbf{a}$ of Euclidean norm $a=\norm{\mathbf{a}}=\sqrt{\mathbf{a}\cdot\mathbf{a}}$.
\section{Elastodynamic Green tensor and its gradient as distributions}
\label{sec:greentens}
\subsection{Navier equation}
The medium is characterized by the elastic tensor $C_{ijkl}$ and the mass density $\rho$. The elastodynamic Green tensor, $G_{ij}$, of the anisotropic Navier equation is defined by
\begin{align}
\label{GF-e}
\left(\delta_{ik}\,\rho\,\partial_{tt}-C_{ijkl}\partial_j\partial_l\right)G_{km}(\rr,t)
=\delta_{im}\,
\delta(t)\delta(\rr)
\end{align}
where $\delta(.)$ denotes the Dirac-delta distribution and $\delta_{ij}$ is Kronecker's symbol. For an isotropic medium,
\begin{align}
\label{eq:celast}
C_{ijkl}=\lambda\delta_{ij}\delta_{kl}+\mu(\delta_{ik}\delta_{jl}+\delta_{il}\delta_{jk}),
\end{align}
where $\lambda$ and $\mu$ are the so-called Lam\'e constants. Substituting Eq.~(\ref{eq:celast}) in Eq.~(\ref{GF-e}), the isotropic Navier equation for the dynamic Green tensor is obtained
\begin{align}
\label{GT-2D-def}
\big[\delta_{ik}\,\rho \, \partial_{tt}- \delta_{ik}\, \mu\, \Delta
-(\lambda+\mu)\, \partial_i \partial_k \big] G_{kj}(\rr,t)=\delta_{ij}\,
\delta(t)\delta(\rr),
\end{align}
where $\Delta$ denotes the Laplacian. In two-dimensional problems, $\rr=(x,y)$. The velocities of the transverse (shear) and longitudinal waves (sometimes called S- and P-waves) are given in terms of the Lam\'e constants as, respectively,
\begin{align}
\label{eq:ctl}
\cT=\sqrt{\mu/\rho},\qquad \cL=\sqrt{(\lambda+2\mu)/\rho}.
\end{align}

\subsection{Anti-plane-strain problem}
In the anti-plane-strain problem, $i=j=z$. If the material is infinitely extended, the Green tensor reduces to $G_{zz}$, the usual Green function of
the two-dimensional scalar wave equation \cite{MORS53,ERIN75,BART89}, solution of
\begin{align}
\label{GT-2D-zz}
\left(\rho\,\partial_{tt}-\mu\Delta\right)G_{zz}=\delta(t)\delta(\rr).
\end{align}
Written as a distribution, its retarded solution reads
\begin{align}
\label{eq:Gzzdist}
G^+_{zz}(\rr,t)=\frac{\theta(t)}{2\pi\mu}\left(t^2-r^2/\cT^2\right)^{-1/2}_+,
\end{align}
where $\theta(t)$ is the Heaviside unit-step function that restricts the solution to positive times. Here, we have denoted with a `plus' superscript the distributional form of the Green function, $G^+_{zz}$, of \emph{real} arguments, to distinguish it from the mere function $G_{zz}(\mathbf{r},t)$, defined as
\begin{align}
\label{eq:Gzzfunc}
G_{zz}(\rr,t)=\frac{1}{2\pi\mu}\left(t^2-r^2/\cT^2\right)^{-1/2},
\end{align}
and whose arguments will be allowed to take on complex values in the next Section.

In Eq.\ (\ref{eq:Gzzdist}), the generalized function $x^\lambda_+$, defined as \cite{SCHW50,GELF64,KANW04}
\begin{align}
\label{x+}
x^\lambda_+=
\left\{
\begin{array}{ll}
\displaystyle{0}\quad &
\displaystyle{\text{for}\ x<0}
\\
\displaystyle{x^\lambda}\quad &
\displaystyle{\text{for}\  x>0}\ , \\
\end{array}
\right.
\end{align}
has been used. The derivative of $x^{1/2}_+$ is given by
\begin{align}
\label{x+_g1}
\big(x^{1/2}_+\big)'=\frac{1}{2} x^{-1/2}_+.
\end{align}
The derivative of $x^{-1/2}_+$ \cite{SCHW50,GELF64,ZEMA65} is the following \emph{pseudo-function}, hereafter denoted by the symbol Pf~:
\begin{align}
\label{x+_g2}
\big(x^{-1/2}_+\big)'=-\frac{1}{2}\Pf x^{-3/2}_+.
\end{align}
Pseudo-functions are particular distributions aimed at regularizing otherwise diverging integrals that involve non locally integrable functions (Ref.\ \cite{SCHW50}, Vol.\ I, pp.\ 38--40; \cite{ZEMA65}, p.\ 17). They encompass Hadamard's finite part and Cauchy's principal value prescriptions, and naturally arise upon differentiating certain distributions, such as the  locally integrable function $x^{-1/2}_+$
in Eq.~(\ref{x+_g2}).

For notational convenience, we introduce the family of functions
\begin{align}
\label{eq:fnu}
f_\nu(r,t)=(t^2-r^2)^{\nu/2},\qquad (\nu\leq 1),
\end{align}
intended to be continued to complex arguments. They are distinguished from the following corresponding distributions of real arguments, denoted with a `plus' superscript:
\begin{align}
\label{eq:fsdef}
f_{\nu}^+(r,t)=\theta(t)\theta(t-r)\Pf (t^2-r^2)^{\nu/2}=\theta(t)\Pf(t^2-r^2)^{\nu/2}_+,
\end{align}
where the symbol $\Pf$ is necessary only when $\nu\leq -2$ (note that the positive-time constraint is included in this definition).

Using Eqs.\ (\ref{x+_g1}) and (\ref{x+_g2}) and the above notations, the gradient of $G^+_{zz}$ reads
\begin{align}
\label{GT-zz-grad}
G^+_{zz,k}(\rr,t)=\frac{\theta(t)}{2\pi\mu}\frac{r_k}{\cT^2}
\Pf\left(t^2-r^2/\cT^2\right)^{-3/2}_{+}=\frac{1}{2\pi\rho}\frac{r_k}{\cT}f^+_{-3}(r,\cT t).
\end{align}

\subsection{Plane-strain problem}
In the plane-strain problem \cite{EASO56,ERIN75}, indices in Eq.~(\ref{GT-2D-def}) take on values $i,j=1,2$, and the retarded distributional solution is the Green tensor ($i,j=1,2$)
\begin{align}
G^+_{ij}(\rr,t)&=\frac{\theta(t)}{2\pi\rho}
\biggl\{\frac{r_i r_j}{r^4}
\Bigl[
t^2\left(t^2-r^2/\cL^2\right)^{-1/2}_{+}
+\left(t^2-r^2/\cL^2\right)^{1/2}_{+}
-t^2\left(t^2-r^2/\cT^2\right)^{-1/2}_{+}\nonumber\\
\label{GT-2D}
&-\left(t^2-r^2/\cT^2\right)^{1/2}_{+}\Bigr]
-\frac{\delta_{ij}}{r^2}
\Bigl[
\left(t^2-r^2/\cL^2\right)^{1/2}_{+}
-t^2 \left(t^2-r^2/\cT^2\right)^{-1/2}_{+}
\Bigr]\biggr\}.
\end{align}
It consists of regular distributions made of locally integrable functions; see Refs.\ \cite{EASO56,ERIN75,KAUS06} for classical (non-distributional) writings of this expression.

The above expression can be re-expressed to emphasize its natural decomposition into spherical and two-dimensional traceless (deviatoric) parts. Introducing the traceless tensor
\begin{align}
\label{eq:tij}
T_{ij}(\bhr)=r^2\partial^2_{ij}\log r=\delta_{ij}-2\,\hr_i\hr_j,
\end{align}
the spherical-deviatoric decomposition of the Green tensor reads, in distributional form, ($i,j=1,2$)
\begin{align}
\label{eq:gija}
G_{ij}^+(\mathbf{r},t)
=\frac{1}{4\pi\rho}\sum_{p=\text{T},\text{L}}\frac{1}{c_p}
\bigg[\delta_{ij}
{}\pm\frac{T_{ij}(\bhr)}{r^2}(2 c_p^2t^2-r^2)\bigg]f_{-1}^+(r,c_p t),
\end{align}
where a `plus' (resp., `minus') sign applies in the sum when $p={\rm T}$ (resp., $p={\rm L}$).  The \emph{function} $G_{ij}(\mathbf{r},t)$ is obtained from this expression by simply removing the $+$ superscript in $f^+_\nu$.

Using Fourier transforms (FT) proves expedient to derive yet another representation, to be employed hereafter. Denoting the FT of $G^+_{ij}(\mathbf{r},t)$ with respect to space variables by $G_{ij}(\mathbf{k},t)$ where $\mathbf{k}$ is the Fourier wavevector, one has in three dimensions (e.g., \cite{PELL10})
\begin{align}
G_{ij}(\mathbf{k},t)
&=\frac{\theta(t)}{\rho k}\left[\frac{1}{\cT}\sin(\cT t k)(\delta_{ij}-\hk_i\hk_j)+\frac{1}{\cL}\sin(\cL t k)\hk_i\hk_j\right]\nonumber\\
\label{eq:gijFT}
&=\frac{\theta(t)}{\rho}\left[\frac{1}{\cT}\frac{\sin(\cT t k)}{k}\delta_{ij}-k_i k_j\sum_{p=\text{T},\text{L}}\frac{(\pm)}{c_p}\frac{\sin(c_p t k)}{k^3}\right],
\end{align}
where $(\pm)$ is a shorthand notation for a factor $(\pm 1)$. Two-dimensional inverses are simply obtained by using the convention that $k=(k_x^2+k_y^2)^{1/2}$ and $k_3=k_z=0$. Employing the integrals
\begin{align}
\int_0^\pi\frac{\d \phi}{2\pi}\,\ee^{\ii x \cos\phi}=J_0(x),\qquad \int_0^\pi\frac{\d \phi}{2\pi}\cos\phi\,\ee^{\ii x \cos\phi}=\ii J_1(x),
\end{align}
where $\cos\phi=\bhk\cdot\bhr$, and $J_\nu$ is the Bessel function, one obtains using Eqs.~(6.671-7) and (3.693-1) in Ref.~\cite{GR}
\begin{subequations}
\begin{align}
\label{eq:ft1}
&\int
\frac{\d^2\!k}{(2\pi)^2}\,\ee^{\ii\mathbf{k}\cdot\mathbf{r}}\,\frac{\sin(c t
  k)}{k}=\frac{1}{2\pi}\int_0^\infty \d k\,J_0(r k)\sin(c t
k)=\frac{1}{2\pi}\left(c^2 t^2-r^2\right)^{-1/2}_+,
\\
\label{eq:ft2}
&\int \frac{\d^2\!k}{(2\pi)^2}\,(\ii k_j)\,\ee^{\ii\mathbf{k}\cdot\mathbf{r}}\,\frac{\sin(c t k)}{k^3}
=-\frac{\hr_j}{2\pi}\int_0^\infty \frac{\d k}{k}J_1(r k)\sin(c t k)
=\frac{\hr_j}{2\pi r}\left[\left(c^2 t^2-r^2\right)_+^{1/2}-c t\right],
\end{align}
\end{subequations}
where in Eq.~(\ref{eq:ft2}) the proportionality to $\hr_j$ of the result was anticipated due of isotropy. Thus, appealing to expressions (\ref{eq:ft1}) and (\ref{eq:ft2}) to invert (\ref{eq:gijFT}) yields $G^+_{ij}(\mathbf{r},t)$ in the following alternative form:
\begin{subequations}
\label{eq:gijdgj}
\begin{align}
\label{eq:gijdgjeq1}
G^+_{ij}(\mathbf{r},t)&=\frac{1}{2\pi\rho}\left\{\frac{1}{\cT}f_{-1}^+(\cT t,r)\delta_{ij}+\frac{1}{2}\left[\partial_i g^+_j(\mathbf{r},t)+\partial_j g^+_i(\mathbf{r},t)\right]\right\}\qquad  (i,j=1,2),\\
g^+_j(\mathbf{r},t)   &=\frac{\hr_j}{r}\sum_{p={\rm T},{\rm L}}\frac{(\pm)}{c_p}[f_{1}^+(c_p t,r)-c_p t]=\frac{\hr_j}{r}\sum_{p={\rm T},{\rm L}}\frac{(\pm)}{c_p}f_{1}^+(c_p t,r).
\end{align}
\end{subequations}
We emphasize that the derivative $\partial_i g_j^+$ is already naturally symmetric with respect to indices $i$ and $j$, so  that the explicit symmetrization in Eq.~(\ref{eq:gijdgjeq1}) is merely a matter of convenience for further use in Sec.\ \ref{sec:applpsp}. Indeed,
\begin{align}
\partial_i g_j^+(\br,t)=\partial_j g_i^+(\br,t)=\frac{1}{2}
\sum_{p={\rm T},{\rm L}}\frac{(\pm)}{c_p}\left[\frac{T_{ij}(\bhr)}{r^2}(2 c_p^2 t^2-r^2)-\delta_{ij}\right]f_{-1}^+(r,c_p t).
\end{align}

To compute the distributional gradient of $G^+_{ij}$, we observe that $\hr_{i,j}=(\delta_{ij}-\hr_i\hr_j)/r$. Consequently
\begin{align}
\label{eq:dtij}
T_{ij,k}(\bhr)=\frac{2}{r}T_{ij}(\bhr)\hr_k-\frac{2}{r}T_{ijk}(\bhr),
\end{align}
where we have introduced the totally symmetric and traceless third-rank tensor
\begin{align}
\label{eq:tijk}
T_{ijk}(\bhr)=\delta_{jk}\hr_i+\delta_{ik}\hr_j+\delta_{ij}\hr_k-4\,\hr_i\hr_j\hr_k.
\end{align}
According to Eq.\ (\ref{eq:gija}), $G^+_{ij}$ is of the type (we omit here for brevity the dependence with respect to time)
\begin{align}
\label{eq:gijtype}
G^+_{ij}(\mathbf{r})=g_s(r)\delta_{ij}+g_d(r)T_{ij}(\bhr),
\end{align}
where $g_s$ and $g_d$ are scalar functions. Using Eq.~(\ref{eq:dtij}), the gradient of Eq.~(\ref{eq:gijtype}) reads
\begin{align}
\label{eq:dgijtype}
G^+_{ij,k}(\mathbf{r})=\left\{g_s'(r)\delta_{ij}+\left[g_d'(r)+\frac{2}{r}g_d(r)\right]T_{ij}(\bhr)\right\}\hr_k
-\frac{2}{r}g_d(r) T_{ijk}(\bhr),
\end{align}
namely,
\begin{align}
\label{eq:gradgijdistr}
G^+_{ij,k}(\mathbf{r},t)\!=\!\frac{1}{4\pi\rho}\sum_{p=\text{T},\text{L}}\frac{1}{c_p}
\Big\{f_{-3}^+(r,c_p t)\big[\delta_{ij}\pm T_{ij}(\bhr)\big]r_k
\mp\frac{2}{r^3}(2 c_p^2t^2-r^2)f_{-1}^+(r,c_p t) T_{ijk}(\bhr)\Big\}.
\end{align}
In expanded form, this reads
\begin{align}
&G^+_{ij,k}(\rr,t)=\frac{\theta(t)}{2\pi\rho}
\bigg\{
\bigg(
\frac{\delta_{ik} r_j+\delta_{jk} r_i }{r^4}-\frac{4 r_i r_j r_k}{r^6}
\bigg)
\Big[
t^2 \big(t^2-r^2/\cL^2\big)^{-1/2}_{+}
+\big(t^2-r^2/\cL^2\big)^{1/2}_{+}
\nonumber\\
&-t^2 \big(t^2-r^2/\cT^2\big)^{-1/2}_{+}
-\big(t^2-r^2/\cT^2\big)^{1/2}_{+}
\Big]
+\frac{2 \delta_{ij} r_k}{r^4}
\Big[
\big(t^2-r^2/\cL^2\big)^{1/2}_{+}
-t^2 \big(t^2-r^2/\cT^2\big)^{-1/2}_{+}
\Big]
\nonumber\\
&+\frac{r_i r_j r_k }{r^4}
\bigg[
\frac{t^2}{\cL^2}\Pf\big(t^2-r^2/\cL^2\big)^{-3/2}_{+}
-\frac{1}{\cL^2}\big(t^2-r^2/\cL^2\big)^{-1/2}_{+}
\nonumber\\
&-\frac{t^2}{\cT^2}\Pf\big(t^2-r^2/\cT^2\big)^{-3/2}_{+}
+\frac{1}{\cT^2}\big(t^2-r^2/\cT^2\big)^{-1/2}_{+}
\bigg]
\nonumber\\
\label{GT-2D-grad}
&+\frac{\delta_{ij}r_k }{r^2}
\bigg[\frac{1}{\cL^2}\big(t^2-r^2/\cL^2\big)^{-1/2}_{+}
+\frac{t^2}{\cT^2}\Pf\big(t^2-r^2/\cT^2\big)^{-3/2}_{+}
\bigg]
\bigg\},
\end{align}
which explicitly features hypersingular kernels interpreted as pseudofunctions, in addition to locally integrable ones.

By means of representation (\ref{eq:gijdgjeq1}) of $G^+_{ij}$, we can organize terms differently in $G^+_{ij,k}$. Using
\begin{align}
&\partial^{2}_{ik}g^+_j(\br,t)
=\partial^{2}_{jk}g^+_i(\br,t)\nonumber\\
&=\sum_{p={\rm T},{\rm L}}\frac{(\pm)}{c_p}
\left\{\frac{1}{2}\left[T_{ij}(\bhr)-
\delta_{ij}\right]r_k f_{-3}^+(r,c_p t)-\frac{1}{r^3}(2 c_p^2 t^2-r^2)f_{-1}^+(r,c_p t)T_{ijk}(\bhr)\right\}
\end{align}
and
\begin{align}
\partial_k f^+_{-1}(r,\cT t)=r_k f^+_{-3}(r,\cT t),
\end{align}
one gets
\begin{align}
&G^+_{ij,k}(\br,t)=\frac{1}{2\pi\rho}\biggl(\frac{r_k}{\cT}f_{-3}(r,\cT t)\delta_{ij}
\nonumber\\
&
+\sum_{p={\rm T},{\rm L}}\frac{(\pm)}{c_p}
\left\{\frac{1}{2}\left[T_{ij}(\bhr)-\delta_{ij}\right]r_k f_{-3}^+(r,c_p t)-\frac{1}{r^3}(2 c_p^2 t^2-r^2)f_{-1}^+(r,c_p t)T_{ijk}(\bhr)\right\}\biggr)\nonumber\\
&=G^+_{zz,k}(\br,t)\delta_{ij}
\nonumber\\
\label{eq:gijkalt}
&
+\frac{1}{2\pi\rho}\sum_{p={\rm T},{\rm L}}\frac{(\pm)}{c_p}
\left\{\frac{1}{2}\left[T_{ij}(\bhr)-
\delta_{ij}\right]r_k f_{-3}^+(r,c_p t)-\frac{1}{r^3}(2 c_p^2 t^2-r^2)f_{-1}^+(r,c_p t)T_{ijk}(\bhr)\right\}.
\end{align}
This expression is employed in Section \ref{sec:applpsp}.

\section{Regularization}
\label{sec:reggf}
To motivate the following developments, we point out that in the problem of Volterra dislocations moving nonuniformly with time-dependent position $\mathbf{s}(t)$ and velocity $\bV(t)=\dot{\mathbf{s}}(t)$, distributional expressions for the material velocity (or the elastic strain field) produced by the dislocations are history-dependent. Namely, they typically involve integrals over past times of the following (or of a related)
type \cite{LAZA14}:
\begin{align}
\label{eq:prototype}
\int_{-\infty}^{t^-} \dd t'\,G^+_{ij,k}(\rr-\mathbf{s}(t'),t-t')\, V_{l}(t'),
\end{align}
where the velocity $\bV(t=-\infty)$ in the remote past is assumed to be a constant (possibly zero), and where the upper boundary $t'=t$ is approached by lower values to
ensure strictly positive time intervals in the derivative of the Green function. The practical necessity for the latter requirement will be illustrated in Sec.\ \ref{sec:apsp}.\footnote{Quite generally, the equal-time value of the time-dependent Green function is defined only as a limit \cite{BART89} (see p.\ 189).
This issue becomes important with the derivatives, and must in general be acknowledged explicitly to carry out calculations properly; see, e.g., Sec.\ 2.2 in \cite{PELL12} for another (related) example where the upper boundary $t'=t$ must be avoided, which is achieved there by means of a specific limiting device. Thus, defining the gradient of the Green tensor almost everywhere is not enough to get a meaningful result in integrals such as (\ref{eq:prototype}). This is another indication ---different from the matter of handling wavefront singularities at finite time intervals--- that the gradient is not integrable in Lebesgue's sense, but is a distribution. We could not find in the literature a proper discussion of this particular point. The present pragmatic treatment, aimed at applications, does not pretend to full mathematical rigor.} The integral expresses the fact that fields at time $t$ arise as sums of field contributions emitted at all `retarded' positions $\mathbf{s}(t')$. An alternative formulation \cite{GURR13} can be derived involving an integral over past positions, which requires computing `retarded times'. For our purpose, the present setting is more straightforward.

The distributional character of $G^+_{ij,k}$ has not been widely acknowledged in the literature, and integrals such as (\ref{eq:prototype}) have sometimes been dismissed as ill-defined. For instance, it has been advocated that derivatives of $G^+_{ij}$ be left \emph{outside} the integral, with expression (\ref{eq:prototype}) written in the form
\begin{align}
\label{eq:prototypeold}
\partial_k\int_{-\infty}^{t^-} \dd t'\,G^+_{ij}(\rr-\mathbf{s}(t'),t-t')\, V_{l}(t').
\end{align}
Although Eq.~(\ref{eq:prototypeold}) is obviously a correct way of carrying out calculations since $G^+_{ij}$ is locally integrable, such precautions prove unnecessary in the framework of distribution theory where writing (\ref{eq:prototype}) is more natural. It must be realized, however, that the equivalent expressions (\ref{eq:prototype}) or (\ref{eq:prototypeold}) are themselves distributions. As will be shown below from Eq.~(\ref{eq:prototype}), this quantity generates Dirac singularities along Mach cones for faster-than-wave motion.

Although being mathematically well-defined in the sense of distributions and therefore free of non-integrable singularities, integrals such as (\ref{eq:prototype}) ---with arbitrarily prescribed motion $\mathbf{s}(t)$--- are inconvenient for numerical evaluation.
For explicit calculations, the Green tensor and its gradient must be regularized. The procedure we call hereafter \emph{isotropic regularization} consists in convoluting $G^+$ by the following isotropic representation of the two-dimensional Dirac distribution (Ref.\ \cite{KANW04}, p.\ 60):
\begin{align}
\delta(\br)=\delta(x)\delta(y)=\lim_{\varepsilon\to 0}\delta^\varepsilon(\br),\qquad\text{with}\qquad
\delta^\varepsilon(\br)=\frac{\varepsilon}{2\pi(r^2+\varepsilon^2)^{3/2}}.
\end{align}
Here $\delta^\varepsilon(\br)$ is a non-singular Dirac-delta sequence with parametric dependence. For $\varepsilon$ finite this corresponds to considering a line source with rotationally-invariant core of radius $\varepsilon$. We denote this convolution product by
\begin{align}
\label{eq:gconvol}
\Giso_{ij}(\br,t)=[G^+_{ij}*\delta^\varepsilon](\br,t)=\int\d^2 r'\,G^+_{ij}(\br-\br',t)\delta^\varepsilon(\br').
\end{align}
It should be noted that other regularizing devices have previously been used for dislocations. For instance, a widely employed model \cite{ESHE49,LOTH92} assumes the dislocation to be extended along its direction of motion and infinitely thin in the direction normal to its glide plane. In the present context, this would consist in convoluting the Green tensor by a one-dimensional Lorentzian core shape ---a procedure known as \emph{harmonic regularization} in the mathematical literature \cite{OBER00}--- with respect to the coordinate along the prescribed direction of motion. Since it breaks isotropy, this procedure is of considerable complexity in the elastodynamic case, for which only the antiplane-strain case (relevant to screw dislocations) has so far been worked out \cite{MARK01}. For that reason this type of regularization is not addressed further in the present work. By contrast, the regularization considered hereafter allows one to tackle both the antiplane-strain and plane-strain (relevant to edge dislocations) cases with only moderate complications.

Using FTs, convolution reduces to multiplying transforms, and carrying out one Fourier inversion. Integrating over the angle first, the FT of
$\delta^\varepsilon$ is simply  (with Eq.~(6.565-3) in Ref.~\cite{GR})
\begin{align}
\delta^\varepsilon(\mathbf{k})=\int\d^2 r\,\delta^\varepsilon(\br) \ee^{-\ii \mathbf{k}\cdot\br}
=\int_0^\infty \d z \,z\,\frac{J_0\big(\varepsilon k z\big)}{(1+z^2)^{3/2}}=\ee^{-\varepsilon k}.
\end{align}
In view of representation (\ref{eq:gijdgj}) of $G^+_{ij}$, and since derivatives can be interchanged with convolution, the expression of $\Giso_{ij}$ follows from multiplying in Eqs.\ (\ref{eq:ft1}) and (\ref{eq:ft2}) the integrands by $\ee^{-\varepsilon k}$, and carrying out next the modified integrals. Thus, for instance,
\begin{align}
\int_0^\infty \d k\, J_0(k r)\sin\big(c t k\big)\ee^{-\varepsilon k}
&=\Im\int_0^\infty \d k\, J_0(k r)\ee^{-(\varepsilon-\ii c t)k}\nonumber\\
&=\Im [(\varepsilon-\ii c t)^2+r^2]^{-1/2}=\Re [(c t+\ii \varepsilon)^2-r^2]^{-1/2},
\end{align}
where the identity
\begin{align}
\label{eq:identsqmz2}
\sqrt{-z^2}=-\ii\sign\big(\Im z\big)z\qquad\text{for}\qquad z\in \mathbb{C}\setminus \mathbb{R}
\end{align}
has been used in the last equality. Likewise, the integral in (\ref{eq:ft2}) is modified as
\begin{align}
\int_0^\infty \frac{\d k}{k}\, J_1(k r)\sin\big(c t k\big)\ee^{-\varepsilon k}
=-\Re\frac{1}{r}[\sqrt{(c t+\ii \varepsilon)^2-r^2}-(c t+\ii \varepsilon)].
\end{align}
\begin{figure}
\centering
\includegraphics[width=16.45cm]{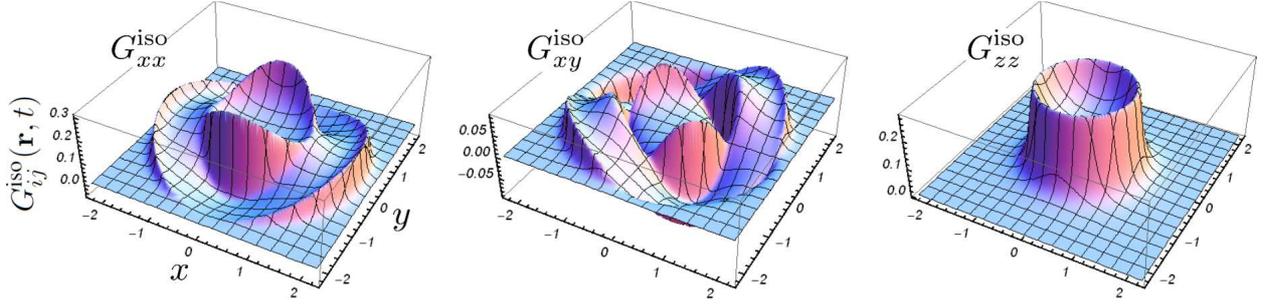}
\caption{\label{fig:fig1}
Independent components of $\Giso_{ij}(\br,t)$ in the $\mathbf{r}=(x,y)$ plane, at $t=1$, for $\cL/\cT=2$ (note the changes of scale on the $G$ axis). Units are such that $\cT=1$ and $\mu=1$, and the regularizing width is $\varepsilon=0.05$. The $yy$ component is the same as the $xx$ one, rotated by $\pi/2$ in the $(x,y)$ plane.}
\end{figure}

Thus, we arrive at the remarkable result that the regularized form $\Giso(\br,t)$ of the \emph{distribution} $G^+_{ij}(\br,t)$ is simply expressed in terms of the \emph{function} $G_{ij}(\br,t)$, continued to complex time, as
\begin{subequations}
\begin{align}
\label{eq:gisoij}
\Giso_{ij}(\br,t)=\theta(t)\Re\left[G_{ij}(\br,t)_{c t\to c t+\ii\varepsilon}\right],
\end{align}
where our notation means that $c_{\rm T} t$ and $c_{\rm L} t$ must be replaced in $G_{ij}(\br,t)$ by $c_{\rm T} t+\ii\varepsilon$ and $c_{\rm L}t+\ii\varepsilon$, respectively. These regularized components are represented in Fig.\ \ref{fig:fig1}, which emphasizes the smoothness of the wavefronts (the value of $\varepsilon$ employed is arbitrary, chosen for best display). It is noted that $G_{zz}^+$ is isotropic in the $(x,y)$ plane, equal to twice the spherical part of the plane-strain Green tensor $G_{ij}^+$~; so that, obviously,
\begin{align}
\label{eq:gisozz}
\Giso_{zz}(\br,t)=\theta(t)\Re\left[G_{zz}(\br,t)_{c t\to c t+\ii\varepsilon}\right].
\end{align}
\end{subequations}
Analytic continuation has long been known as a method of defining distributions (\cite{GELF64} and \cite{KANW04}, p.\ 159). Bremermann's approach \cite{BREM61,BREM65,CARM89} consists in defining distributions as the boundary values of analytic functions on the real axis, the main variable of integration being extended into a complex quantity. Here, in a multi-dimensional context involving space and time variables, we make a definite connection between the latter approach and the particular analytic continuation introduced here, motivated by the issue of handling a line source with finite-size core.

\subsection{Regularized gradient of the Green tensor}
The regularized form of any derivative is obtained from straightforwardly differentiating its regularized primitive (e.g., \cite{KANW04} p.\ 80), which here follows from the commutativity property between convolution and differentiation.  Thus, the expression of $\Giso_{ij,k}$ results from differentiating Eqs.~(\ref{eq:gisoij}) and (\ref{eq:gisozz}), where the functions $G_{zz}$ and $G_{ij}$  are obtained from Eqs.~(\ref{eq:Gzzfunc}) and (\ref{eq:gija}) by replacing the distribution $f_{-1}^+$ by the function $f_{-1}$.

Some care must be exercised when differentiating $f_{-1}(r,ct+\ii\varepsilon)=[(ct+\ii\varepsilon)^2-r^2]^{-1/2}$.
\ref{sec:dcp} emphasizes the fact that whenever the imaginary part of the argument of the square root (computed as a principal determination) depends on the variable with respect to which differentiation is carried out, taking a distributional derivative generates a Dirac term supported by the branch cut of the square root. In the present case however, the group under the square root has imaginary part $\Im[(ct+\ii\varepsilon)^2-r^2]=2\varepsilon c t$. As this does not depend on $\br$, Eq.\ (\ref{eq:zmonehalf}) tells us that, trivially,
\begin{align}
\label{eq:gradiso}
\frac{\partial}{\partial r_k}\frac{1}{\sqrt{(c t+\ii\varepsilon)^2-r^2}}
=\frac{r_k}{[(c t+\ii\varepsilon)^2-r^2]^{3/2}},
\end{align}
so that taking spatial derivatives of such algebraic terms is straightforward with the regularization considered.

It follows that the regularized gradient is simply
\begin{align}
\Giso_{ij,k}(\br,t)=\theta(t)\Re \left[G_{ij,k}(\br,t)_{c t\to c t+\ii\varepsilon}\right],
\end{align}
where $G_{ij,k}(\br,t)$ is the \emph{function} obtained from the distributional expressions of the gradients (\ref{GT-zz-grad}) (anti-plane-strain) or (\ref{eq:gradgijdistr}) (plane-strain), by replacing the distributions $f^+_{-1}$ and $f^+_{-3}$ by the functions $f_{-1}$ and $f_{-3}$.

\begin{figure}[1ht]
\centering
\includegraphics[width=8cm]{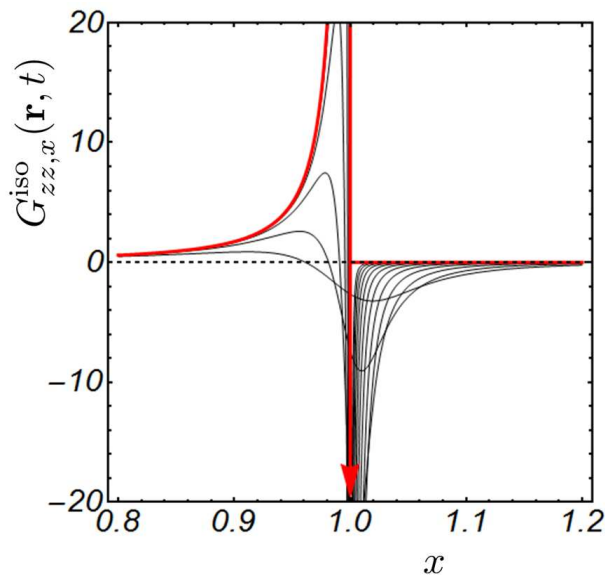}
\caption{
\label{fig:fig2} Component $\Giso_{zz,x}(\br,t)$ near the wavefront $\cT t^2-r^2=0$, at $\br=(x,y=0)$ and for $t=1$, vs.\ $x$, for $\varepsilon=2^{-k}$, with $k=4,5,\ldots,15$ (solid black). For $x<1$, the plots cannot be discriminated when $k\geq 8$ (i.e., $\varepsilon<4.\,10^{-3}$). Units are such that $\cT=1$ and $\mu=1$. Red: representation of the limiting distribution when $\varepsilon\to 0$ (see text).}
\end{figure}
Fig.\ \ref{fig:fig2} illustrates on $\Giso_{zz,x}$ the way isotropic regularization handles the finite-part prescription of $\Pf(c^2 t^2-r^2)_+^{-3/2}$ in the distributional derivatives $G^+_{ij,k}(\br,t)$, as $\varepsilon\to 0$. A negative peak of width proportional to $\varepsilon$ is created on the side $r>c t$ of the singular wavefront $c^2 t^2-r^2=0$. This peak has a sign opposite to that of the peak on the other side $r<c t$. The heights of both peaks are proportional to $\varepsilon^{-3/2}$, so that the negative peak compensates for the non-integrable part of $G_{ij,k}$. In the figure we attempted to represent the limiting distribution $(2\pi\rho\cT)^{-1}\Pf(\cT^2 t^2-r^2)^{-3/2}_+ x$, symbolizing its finite-part prescription by a downward vertical arrow. The imperfect character of this representation illustrates the well-known fact that distributions are linear functionals, often unable to deliver useful numbers unless applied to test functions. The only truly meaningful plots are those drawn with $\varepsilon$ finite.

\section{A case of nonuniformly-moving line source}
\label{sec:movingline}
\subsection{Fundamental integral, and notations}
We apply the regularized Green functions obtained in the previous Section to the regularization of the following fundamental definite integral, defined for $t>0$:
\begin{align}
\label{eq:iijkrtdef}
I_{ijk}(\br,t)&=\int_0^{t^-} G^+_{ij,k}(\br-\bV t',t-t')\dd t'
=\int_{0^+}^t G^+_{ij,k}(\br-\bV (t-\tau),\tau)\dd \tau,\nonumber\\
&\hspace{5cm}(i=j=z,\text{ or }i,j=1,2; k=1,2),
\end{align}
where the change of variables $\tau=t-t'$ has been used. This integral arises as a particular instance of Eq.~(\ref{eq:prototype}) with $\Bs(t')=\bV t'$, in the problem of a dislocation (or a line force) at rest at negative times and instantaneously accelerated to constant velocity $\bV$ at $t=0$ \cite{MARK01,PELL12}.
As the acceleration is instantaneous, this is the simplest prototypical case of nonuniform motion. The results obtained below are important elements of its full solution \cite{LAZA14}. The velocity factor that enters the integrand of integral (\ref{eq:prototype}) being a constant in the case considered, it has been factored out and omitted in the above definition of $I_{ijk}(\br,t)$ (this is the reason why, e.g., in Eq.\ (\ref{eq:vzer}) of in Fig.\ \ref{fig:fig3} below, a nonvanishing field pattern involving an acceleration wave will be obtained for zero velocity).

For definiteness we focus on the time interval $\tau \in(0^+,t),$ but the result will essentially take the form of a difference between values of \emph{indefinite} integrals ---denoted by $I_k(\br,\tau)$ (anti-plane-strain case) or $J_{ijk}(\br,\tau)$ (plane strain case)--- at the boundaries $\tau=t$ and $\tau=0^+$. Thanks to these indefinite integrals, to be derived below for any orientation of $\bV$, more general time intervals could be considered, which is useful for numerical purposes (see Introduction).

Let $c$ be a placeholder for either $\cT$ or $\cL$.  The following notations and quantities are employed hereafter:
\begin{subequations}
\begin{align}
\bbeta&=\bV/c,\qquad
\beta =|\!|\bbeta|\!|,\qquad
\gamma=1/\sqrt{1-\beta^2},\\
\bR(\tau)&=\br+\bV \tau,\\
S(\tau)&=\sqrt{c^2\tau^2-R(\tau)^2},\\
D(\tau)&=c\tau-\bbeta\cdot\bR(\tau),\\
\label{eq:apm}
\Apm_{ij}(\bbeta)&=(1-\beta^2)(\delta_{ij}-\hn_i\hn_j)\pm\hn_i\hn_j,
\end{align}
\end{subequations}
where here and in the rest of the paper the unit vector $\bhn=\bhV=\bhbeta$ indicates the direction of motion, and $\tau$ is a time variable. With respect to this direction, we furthermore introduce the following decomposition of the position vector (where $r_\perp\geq 0$ and $r_\parallel$ can be of any sign):
\begin{align}
\label{eq:parperp}
r_\parallel=\br\cdot\bhn,\qquad
\br^\perp=\br-r_\parallel\bhn,\qquad
r_\perp=\norm{\br^\perp},\qquad
r_\parallel^2+r_\perp^2=r^2.
\end{align}

\subsection{Anti-plane-strain problem}
\label{sec:apsp}
In this Section we let $c=\cT$ to simplify notations, the longitudinal wave speed $\cL$ being irrelevant. Going to the co-moving frame by changing $\br$ into $\br+\bV t$, the anti-plane problem consists in obtaining the regularized forms of
\begin{align}
I_{zzk}(\br+\bV t,t)=\int_{0^+}^t G^+_{zz,k}(\br+\bV \tau,\tau)\dd\tau.
\end{align}

With $\mu=\rho c^2$, the regularized form of this integral is
\begin{align}
I^{\text{iso}}_{zzk}(\br+\bV t,t)
&=\int_{0^+}^t \Giso_{zz,k}(\br+\bV \tau,\tau)\dd\tau=\frac{1}{2\pi\rho c}\Re\int_{0^+}^t \frac{R_k(\tau)}{[(c\tau+\ii\varepsilon)^2-R^2(\tau)]^{3/2}}\dd\tau\nonumber\\
\label{eq:iisozzk}
&=\frac{1}{2\pi\mu}\Re\left[c\int_{0^++\ii\varepsilon/c}^{t+\ii\varepsilon/c} S(\tau)^{-3}R_k(\tau)\dd\tau\right]_{\br\to\br-\ii\varepsilon\bbeta}.
\end{align}
where the prescription $\br\to\br-\ii\varepsilon\bbeta$ is a consequence of having shifted the integration path by a value $+\ii\varepsilon/c$. It should be noted that the combination of the shifts $c\tau\to c\tau-\ii\varepsilon$
and $\br\to\br+\ii\varepsilon\bbeta$ leaves $\bR=\br+\bV\tau$ invariant. Because the imaginary part under the square root $S(\tau)$ is strictly positive on this path [see remark preceding Eq.\ (\ref{eq:gradiso})], the branch cut of the square root is never crossed. Thus, the integral can be computed as the difference of boundary values of the following indefinite integral, easily verified by differentiation:
\begin{align}
\label{eq:indint1}
I_k(\br,\tau)=c\int \dd\tau\, S(\tau)^{-3}(r_k+V_k\tau)
=\frac{(r_k\beta_p-\beta_k r_p)(r_p+V_p\tau)-c\tau\,r_k}{\left(\br\cdot\sfAplus\cdot\br\right)S(\tau)}.
\end{align}
It follows that
\begin{align}
\label{eq:Iisozzk}
I^{\text{iso}}_{zzk}(\br+\bV t,t)
=\frac{1}{2\pi\mu}\Re\left[I_k(\br-\ii\varepsilon\bbeta,t+\ii\varepsilon/c)
-I_k(\br-\ii\varepsilon\bbeta,0^++\ii\varepsilon/c)\right].
\end{align}
This case is sufficiently simple that, using the `parallel' and `perpendicular' notation [Eq.\ (\ref{eq:parperp})], and
\begin{align}
(\br-\ii\varepsilon\bbeta)\cdot\sfAplus\cdot(\br-\ii\varepsilon\bbeta)
= \left[(1-\beta^2)r_\perp^2+(r_\parallel-\ii\varepsilon \beta)^2\right],
\end{align}
we can give the result explicitly as
\begin{align}
I^{\text{iso}}_{zzk}(\br+\bV t,t)&=\frac{1}{2\pi\mu}\Re\biggl\{
\frac{1}{(1-\beta^2)r_\perp^2+(r_\parallel-\ii\varepsilon\beta)^2}\times\nonumber\\
\label{eq:Iisozzexpl}
&\hspace{-5em}\times\biggl[
\frac{(r_k\beta_p-\beta_k r_p)(r_p+V_p t)-(ct+\ii\varepsilon)(r_k-\ii\varepsilon\beta_k)}{
\sqrt{(ct+\ii\varepsilon)^2-|\!|\br+\bV t|\!|^2}}
-\frac{(r_k\beta_p-\beta_k r_p)r_p-\ii\varepsilon(r_k-\ii\varepsilon\beta_k)}{\ii\sqrt{r^2+\varepsilon^2}}
\biggr]
\biggr\}.
\end{align}
In the rightmost term, the correct sign of the square root $\sqrt{(0^++\ii\varepsilon)^2-r^2}$ $=\sqrt{-(r^2+\varepsilon^2)+\ii 0^+}$ $=\ii\sqrt{r^2+\varepsilon^2}$ stems from the prescription $0^+$ employed
in Eq.~(\ref{eq:Iisozzk}). This is the reason why approaching the upper-time boundary $t$ as a limit, in time integrals such as Eq.~(\ref{eq:iijkrtdef}), is required to obtain a well-defined result. This limit must be taken with $\varepsilon$ finite, before possibly letting $\varepsilon\to 0$.

The quantity $I^{\text{iso}}_{zzk}(\br,t)$ follows from going back to the reference frame by substituting $\br$ by $\br-\bV t$ in this expression. Thus,
\begin{align}
\label{eq:izziso}
I^{\text{iso}}_{zzk}(\br,t)&=\frac{1}{2\pi\mu}\Re\biggl(
\frac{1}{(1-\beta^2)r_\perp^2+[r_\parallel-(c t+\ii\varepsilon)\beta]^2}\times\\
&\hspace{-5em}\times\biggl\{
\frac{(r_k\beta_p-\beta_k r_p)r_p-(ct+\ii\varepsilon)[r_k-(c t+\ii\varepsilon)\beta_k]}{
\sqrt{(ct+\ii\varepsilon)^2-r^2}}
+\ii\frac{(r_k\beta_p-\beta_k r_p)(r_p-c t\beta_p)-\ii\varepsilon[r_k-(c t+\ii\varepsilon)\beta_k]}{\sqrt{(\smash{r_\parallel} -c t\beta)^2+\smash{r_\perp}^2+\varepsilon^2}}
\biggr\}
\biggr).\nonumber
\end{align}

\begin{figure}[1ht]
\centering
\includegraphics[width=16.45cm]{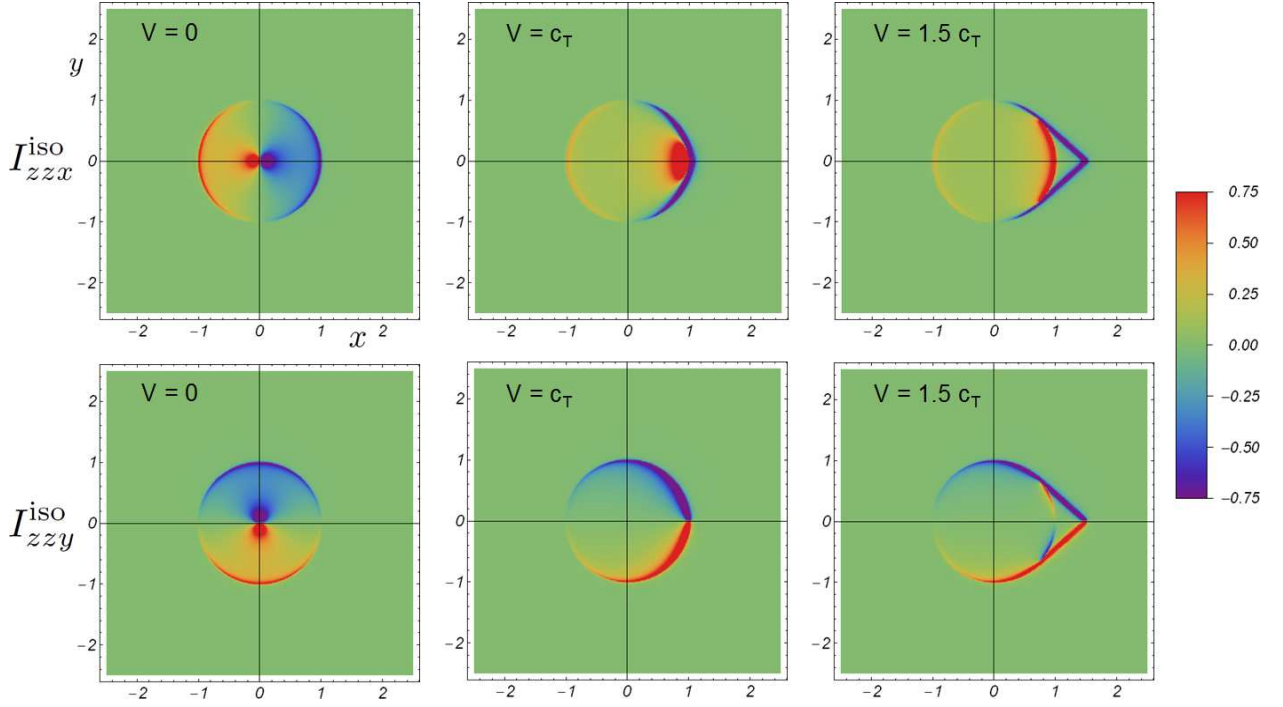}
\caption{\label{fig:fig3}
Anti-plane strain problem, with $\varepsilon=0.01$. Components $I^{\text{iso}}_{zzx}(\br,t)$ and $I^{\text{iso}}_{zzy}(\br,t)$ at time $t=1$ vs.\ position $\br=(x,y)$ for a line source moving in the positive direction along the $Ox$ axis, at constant velocity $V$ (as indicated) after having been instantaneously accelerated from rest at the origin of coordinates at $t=0$ (see text for the case $V=0$).  Units are such that $\cT=1$ and $\mu=1$. For better display fields have been thresholded as indicated in the bar legend.}
\end{figure}
Fig.~\ref{fig:fig3} represents $I^{\text{iso}}_{zzx}(\br,t)$ and  $I^{\text{iso}}_{zzy}(\br,t)$, with $\varepsilon$ small enough to highlight the details of the field structure. The circular shear wave, centered on the origin, is emitted at $t=0$, and expands linearly with time at wavespeed $\cT$. The figure illustrates the fact that via analytic continuation, the proposed regularization handles faster-than-wave source motion at no additional price. Waves patterns are very different depending on the component considered: the source acts as a dipole oriented along $Ox$ in the $zzx$ component, and as one oriented along $Oy$ in the $zzy$ component. As a consequence, the Mach cone for $|V|>\cT$ in the $zzy$ component has branches of opposite signs on either side of the $Ox$ axis, and the leading part of the circular wave experiences a change of sign as well.

\subsection{Limiting distribution in the anti-plane-strain problem, and Mach-cone analysis}
In this section, we show how to compute the limit of expression (\ref{eq:izziso}) as $\varepsilon\to 0^+$, and investigate some of its properties. This limit can be considered a \emph{definition} of the corresponding distribution $I_{zzk}^+(\br,t)$. As a warm-up, we consider first the case of zero velocity, for which Eq.\ (\ref{eq:izziso}) reduces to
\begin{align}
\label{eq:vzer}
I^{\text{iso}}_{zzk}(\br,t)|_{\bV=\mathbf{0}}&=\frac{1}{2\pi\mu}\frac{r_k}{r^2}
\biggl[\frac{\varepsilon}{\sqrt{r^2+\varepsilon^2}}-\Re\frac{(ct+\ii\varepsilon)}{\sqrt{(ct+\ii\varepsilon)^2-r^2}}
\biggr].
\end{align}
This expression is nonsingular as $r\to 0$. Obviously,
\begin{align}
\label{eq:sqroots}
\lim_{\varepsilon\to 0}\frac{1}{\sqrt{(ct+\ii\varepsilon)^2-r^2}}=(c^2t^2-r^2)^{-1/2}_+ -\ii (r^2-c^2t^2)_+^{-1/2}.
\end{align}
It follows that in the limit,
\begin{align}
\label{eq:vzerlim}
I^+_{zzk}(\br,t)|_{\bV=\mathbf{0}}&=-\frac{ct}{2\pi\mu}(c^2t^2-r^2)_+^{-1/2}\frac{r_k}{r^2},
\end{align}
which is locally integrable in two dimensions, and therefore defines a distribution. Indeed, for any function $f(\br)$ finite at $\br=\mathbf{0}$ and vanishing fast enough at infinity, the following integral (where $\varphi$ is the polar angle) is well-defined:
\begin{align}
\int \dd^2r\, \frac{r_k}{r^2}f(\br)=\int_0^\infty \dd r\int_0^{2\pi}\dd\varphi\,\hat{r}_k f(\br).
\end{align}
Although this is not needed in the above example, we remark that a convenient way of evaluating the function $1/r$ as a distribution (and of obtaining its derivatives) is to write it $\lim_{\varepsilon\to 0}\theta(r-\varepsilon)/r$ (\cite{KANW04}, p.\ 135).

We now turn to Eq.\ (\ref{eq:izziso}), assuming that $\bV\not=\mathbf{0}$, so that $\beta>0$. We must also assume that $\beta\not=1$, the special difficulty presented by the value $\beta=1$ being postponed at the end of this section. To compute the limit, we start off by invoking the Sokhotski--Plemelj formula (e.g., Ref.\ \cite{KANW04}, p.\ 27)
\begin{align}
\label{eq:plemelj}
\frac{1}{x\pm \ii 0^+}=\pv\frac{1}{x}\mp\ii\pi\delta(x),
\end{align}
where `$\pv$' denotes a principal value. Thus, the overall factor in Eq.~(\ref{eq:izziso}) becomes
\begin{align}
&\lim_{\varepsilon\to 0}\frac{1}{(1-\beta^2)r_\perp^2+[r_\parallel-(c t+\ii\varepsilon)\beta]^2}
=\lim_{\varepsilon\to 0}\frac{1}{(1-\beta^2)r_\perp^2+(r_\parallel-c t\beta)^2-2\ii\varepsilon\beta(r_\parallel-c t\beta)}\nonumber\\
\label{eq:limits}
&=\pv\frac{1}{(1-\beta^2)r_\perp^2+(r_\parallel-c t\beta)^2}+\ii\pi\sign(r_\parallel-c t\beta)\delta\left((1-\beta^2)r_\perp^2+(r_\parallel-c t\beta)^2\right).
\end{align}

Next, the leftmost square root in the term within braces in Eq.~(\ref{eq:izziso}) is addressed by means of Eq.\ (\ref{eq:sqroots}). Besides, because it is of positive argument, the rightmost square root in Eq.\ (\ref{eq:izziso}) is nonsingular; it defines a locally integrable term in the limit. Finally, the harmless powers of $\varepsilon$ in the numerators can be set to zero right away. Thus,
introducing
\begin{subequations}
\begin{align}
\Delta(\br,t)&=(\br-ct\bbeta)\cdot\sfAplus\cdot(\br-ct\bbeta)\nonumber\\
\label{eq:cons1}
&=(r_\parallel-c t\beta)^2-(\beta^2-1)r_\perp^2\\
\label{eq:cons2}
&=(\beta r_\parallel-c t)^2-(\beta^2-1)(r^2-c^2 t^2),
\end{align}
\end{subequations}
one arrives at
\begin{align}
\label{eq:izzktmp}
I^+_{zzk}(\br,t)&=\frac{1}{2\pi\mu}\Re\biggl(
\left[\pv\frac{1}{\Delta(\br,t)}+\ii\pi\sign(r_\parallel-c t\beta)\,\delta\bigl(\Delta(\br,t)\bigr)\right]\\
&\hspace{-5em}\times\biggl\{
\left[(r_k\beta_p-\beta_k r_p)r_p-ct(r_k-c t\beta_k)\right]\left[(c^2t^2-r^2)^{-\frac{1}{2}}_+ -\ii (r^2-c^2t^2)_+^{-\frac{1}{2}}\right]
+\ii\frac{(r_k\beta_p-\beta_k r_p)(r_p-c t\beta_p)}{\sqrt{(\smash{r_\parallel}-c t\beta)^2+\smash{r_\perp}^2}}
\biggr\}
\biggr).\nonumber
\end{align}

\begin{figure}[!ht]
\centering
\includegraphics[width=10cm]{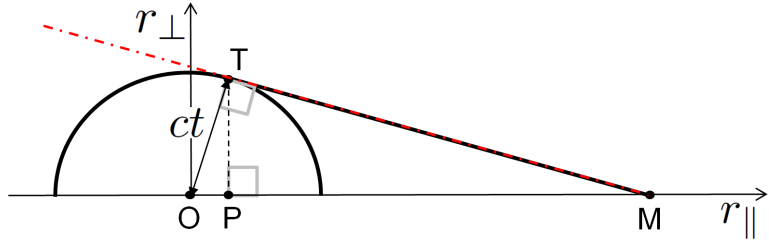}
\caption{
\label{fig:fig4}
Locus of the Mach cone for $\beta>1$.  By definition [Eq.\ (\ref{eq:parperp})], the O$r_\parallel$ axis is oriented along $\bV$ and the transverse O$r_\perp$ coordinate axis is positive, so that only half of the wave pattern is represented. The Mach cone is tangent at point T to the circular wave of radius OT${}=ct$, emitted at $t=0$ by the source point M. Given the traveling distance OM${}=ct\beta$, the geometric construction shows that OP${}=ct/\beta$. Points on the Mach cone are such that $ct/\beta<r_\parallel<ct\beta$ and $r>ct$. Dot-dashed (red): supporting line of the Mach cone.
}
\end{figure}
The function $\Delta(\br,t)$ vanishes only at the source position $\br=\bV t=ct\bbeta$ if $\beta\leq 1$. On the other hand, for faster-than-wave motion with $\beta>1$ it vanishes for $r_\perp=|r_\parallel-ct\beta|/\sqrt{\beta^2-1}$, namely, on the supporting line of the Mach cone drawn in Fig.\ \ref{fig:fig4}, which encompasses the source position (because the coordinate system employed is such that $r_\perp\geq 0$, only one half of the physical wavefront pattern is represented). Thus, Mach-cone contributions for $\beta>1$ in Eq.\ (\ref{eq:izzktmp}) are concentrated in the term $\delta\bigl(\Delta(\br,t)\bigr)$. We thus retrieve the well-known fact that the Mach cone generated by a point dislocation involves the Dirac distribution \cite{CALL80}.

We are now in position to simplify Eq.~(\ref{eq:izzktmp}). Under the Dirac constraint, expressions (\ref{eq:cons1}) and (\ref{eq:cons2}) are employed to express both $r_\perp^2$ and $\sqrt{r^2-c^2 t^2}$ in terms of $r_\parallel$. Extracting the real part out of Eq.~(\ref{eq:izzktmp}), this allows one to turn the terms with square roots in the prefactor of $\delta\bigl(\Delta(\br,t)\bigr)$, into rational fractions involving absolute values. Factoring signs out of the latter, further simplifications lead to
\begin{align}
\label{eq:izzkdistr}
I^+_{zzk}(\br,t)&=\frac{1}{\mu}\left[C_k(\br,t)+M_k(\br,t)\right],
\end{align}
where the circular wave $\mathbf{C}$ and Mach-cone $\mathbf{M}$ vector contributions have, respectively, components
\begin{subequations}
\begin{align}
\label{eq:cwt}
C_k(\br,t)&=\frac{1}{2\pi}(c^2t^2-r^2)^{-\frac{1}{2}}_+
\pv\frac{(r_k\beta_p-\beta_k r_p)r_p-ct(r_k-c t\beta_k)}{\Delta(\br,t)},\\
\label{eq:mct}
M_k(\br,t)&=-\Theta(r_\parallel,t)\,(\beta^2-1)_+^{-\frac{1}{2}}[(r_k-c t\beta_k)-\beta_p(\beta_p r_k- r_p\beta_k)]\delta\bigl(\Delta(\br,t)\bigr).
\end{align}
\end{subequations}
In definition (\ref{eq:mct}), the quantity
\begin{align}
\Theta(r_\parallel,t)=\frac{1}{2}\left[1-\sign(\beta r_\parallel-c t)\sign(r_\parallel-c t\beta)\right]
\end{align}
is a characteristic function equal to $1$ in the interval $r_\parallel\in (ct/\beta,ct\beta)$, and to 0 otherwise. It thus restricts the Mach cone to its physical range (segment TM in Fig.\ \ref{fig:fig4}).

\begin{figure}[!ht]
\centering
\includegraphics[width=9cm]{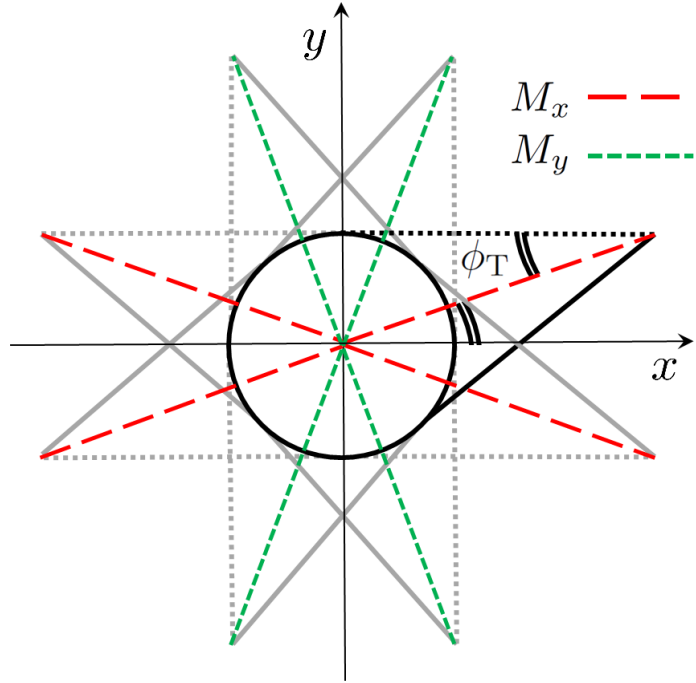}
\caption{
\label{fig:fig5} Particular velocity directions for which the Mach-cone branch parallel to one of the Cartesian coordinate axes is extinguished (dotted) in components $I^+_{zzk}$. Long-dashed, red: directions relevant to component $M_x$; short-dashed, green: \textit{idem}, for $M_y$. The angle $\phi_{\rm T}$ is the half-cone aperture.}
\end{figure}
Assuming $\beta>1$, further insight into the Mach-cone part $\bM(\br,t)$ is obtained by using basis orthogonal vectors $\bhe_{x,y}$ and corresponding coordinates $x$ and $y$. Let moreover $\bbeta=\beta(\cos\phi\,\bhe_x+\sin\phi\,\bhe_y)$, so that $r_\parallel(x,y)=x\cos\phi+y\sin\phi$. The quantity $\Delta(\br,t)$ admits the factorization
\begin{subequations}
\begin{align}
\label{eq:deltfact}
\Delta(\br,t)&=\Delta_+(\br,t)\Delta_-(\br,t),\\
\Delta_\pm(\br,t)&=\left(\cos\phi\pm\sqrt{\beta^2-1}\sin\phi\right)x+\left(\sin\phi\mp\sqrt{\beta^2-1}\cos\phi\right)y-c t \beta.
\end{align}
\end{subequations}
Considering $t$ as the main variable in $\delta\bigl(\Delta(\br,t)\bigr)$ puts the variables $x$ and $y$ on an equal footing, which is most desirable here. Let thus $t_\pm(\br)$ be the respective solutions of $\Delta_\pm(\br,t)=0$. According to a well-known property of the Dirac distribution (e.g., \cite{KANW04} p.\ 49), we have from (\ref{eq:deltfact})
\begin{align}
\delta\left(\Delta(\br,t)\right)=\sum_{s=\pm 1}\frac{\delta\left(\Delta_s(\br,t)\right)}{|\Delta_{-s}(\br,t_s(\br))|}.
\end{align}
Inserting this expression into (\ref{eq:mct}) yields
\begin{align}
\label{eq:mexpl1}
\bM(\br,t)
&=-\frac{1}{2}\sign\left(y\cos\phi-x\sin\phi\right)\times\\
&\times\sum_{s=\pm 1}\left[
\left(\frac{s\,\cos\phi}{\sqrt{\beta^2-1}}+\sin\phi\right)\bhe_x
+\left(-\cos\phi+s\,\frac{\sin\phi}{\sqrt{\beta^2-1}}\right)\bhe_y\right]
\Theta\bigl(r_\parallel,t_s(\br)\bigr)\delta\bigl(\Delta_s(\br,t)\bigr).\nonumber
\end{align}
This expression shows that one of the two branches of the Mach cone component is ``extinguished'' whenever $\phi=\pm\phi_{\rm T}$ or $\phi=\pi\pm\phi_{\rm T}$ for $M_x$, and for $\phi=\pi/2\pm\phi_{\rm T}$ or $\phi=-\pi/2\pm\phi_{\rm T}$ for $M_y$, where $\phi_{\rm T}=\arctan 1/\sqrt{\beta-1}$ is the half-opening angle of the cone. For each of these angles, the extinguished branch lies along the $Ox$ axis for $M_x$, and along the $Oy$ axis for $M_y$ (Fig.\ \ref{fig:fig5}). This accidental extinction, to be further illustrated in the next section, is a mere consequence of the tensor character of the integral under consideration. However, it is impossible to turn all components off simultaneously, which is physically obvious.

As to $\mathbf{C}(\br,t)$, if $\beta<1$ the principal-value prescription in Eq.\ (\ref{eq:cwt}) handles the singularity at the source point, which then lies within the circular wave boundary. If on the contrary $\beta>1$, this unique singularity splits into two singular points that stand at the intersection of the circular wavefront and of the Mach cone lines (tangency points represented by $T$ in Fig.\ \ref{fig:fig4}). Their investigation can be carried out straightforwardly by decomposing the overall rational fraction into partial ones, using the factorization formula (\ref{eq:deltfact}). As these field singularities are well known \cite{CALL80,MARK08}, the analysis will not be pursued further.

Before closing the section, we return for completeness to the marginal case $\beta=1$. Then, Eq.\ (\ref{eq:limits}) is meaningless, because we face the problem of a double root in the denominator. The case must be addressed by the following generalization (\cite{GELF64}, p.\ 60) of the Sokhotski-Plemelj formula:
\begin{align}
\frac{1}{(x\pm \ii 0^+)^2}=\Pf \frac{1}{x^2}\pm\ii\pi\delta'(x),
\end{align}
which stems from differentiating (\ref{eq:plemelj}). Thus for $\beta=1$ equation (\ref{eq:limits}) becomes
\begin{align}
\label{eq:limits2}
&\lim_{\varepsilon\to 0}\frac{1}{[r_\parallel-(c t+\ii\varepsilon)]^2}
=\Pf\frac{1}{(r_\parallel-c t)^2}-\ii\pi\delta'\!\left(r_\parallel-c t\right),
\end{align}
The analysis can then be completed as above, separating $I^+_{zzk}(\br,t)$ into ``circular wave'' and ``cone'' parts. By (\ref{eq:limits2}), the cone part is formally proportional to $\delta'(r_\parallel-c t)=-c^{-2}\delta'(t-r_\parallel/c)$. It vanishes, as must be, due to the identity (\cite{KANW04}, p.\ 36):
\begin{align}
f(t)\delta'(t-t_0)=f(t_0)\delta'(t-t_0)-f'(t_0)\delta(t-t_0),
\end{align}
where $f$ is some differentiable function at $t=t_0$. Applying this identity with $t_0=r_\parallel/c$ to the prefactor of $\delta'(t-r_\parallel/c)$ ---considered as a function of time--- shows that the whole term cancels out. The surviving circular-wave contribution reads
\begin{align}
I_{zzk}^+(\br,t)&=\frac{1}{2\pi\mu}(c^2t^2-r^2)^{-\frac{1}{2}}_+
\Pf\frac{(r_k\beta_p-\beta_k r_p)r_p-ct(r_k-c t\beta_k)}{(r_\parallel-c t)^2}\qquad (\beta=1).
\end{align}
We note that this expression is nonsingular even when $r_\perp=0$. Indeed, the numerator of the fraction reduces then to
\begin{align}
(r_k\beta_p-\beta_k r_p)r_p-ct(r_k-c t\beta_k)=(r_\parallel-ct)\left[r_k-(r_\parallel+ct)\widehat{\beta}_k\right]\qquad (\beta=1, r_\perp=0).
\end{align}
Moreover, $\left(c^2t^2-r^2\right)^{1/2}=(c^2t^2-r_\parallel^2)^{1/2}$, so that the distribution behaves as $|ct-r_\parallel|^{-3/2}$ which remains regularized by the $\Pf$ prescription.

\subsection{Plane-strain problem}
\label{sec:applpsp}
Evaluating the plane-strain integral
\begin{align}
\label{eq:Iijk}
I_{ijk}(\br+\bV t,t)=\int_{0^+}^t G^+_{ij,k}(\br+\bV \tau,\tau)\dd \tau,\qquad i,j,k=1,2,
\end{align}
is notably more complicated. We employ the same method as for the anti-plane-strain case, our efforts being directed towards obtaining the key indefinite integral that allows for a direct solution.

The form (\ref{eq:gijdgjeq1}) of $G_{ij}^+$ and the associated representation (\ref{eq:gijkalt}) of its gradient carry over \textit{mutatis mutandis} to a similar representation for $\Giso_{ij,k}$. In terms of the latter the integral we need to compute is:
\begin{align}
&I^{\text{iso}}_{ijk}(\br+\bV t,t)
=\int_{0^+}^t\Giso_{ij,k}(\br+\bV \tau,\tau)\dd\tau\nonumber\\
&=I^{\text{iso}}_{zzk}(\br+\bV t,t)\delta_{ij}\nonumber\\
\label{eq:iisoijkdef}
&+\frac{1}{2\pi\rho}\Re\sum_{p={\rm T},{\rm L}}\frac{(\pm)}{c_p}
\int_{0^++\ii\varepsilon/c_p}^{t+\ii\varepsilon/c_p}
\frac{\dd\tau}{2}\left\{\partial^2_{ik}\left[S(\tau)\frac{R_j(\tau)}{R(\tau)^2}\right]_{c\to c_p}+(i\leftrightarrow j)\right\}_{\br\to\br-\ii\varepsilon\bbeta^{(p)}},
\end{align}
where $\bbeta^{(p)}=\bV/c_p$, and where $I^{\text{iso}}_{zzk}(\br+\bV t,t)$ is given by Eq.\ (\ref{eq:Iisozzk}). By the same arguments as in the previous Section, the above integral follows from  the indefinite integral
\begin{subequations}
\label{eq:jijkgen}
\begin{align}
\label{eq:jijkgen1}
J_{ijk}(\br,\tau)
&=c\int\frac{\dd\tau}{2}\left\{\partial^2_{ik}\left[S(\tau)\frac{R_j(\tau)}{R(\tau)^2}\right]+(i\leftrightarrow j)\right\}\\
\label{eq:jijkgen2}
&=c\int\dd \tau\,\left\{\frac{1}{2}\frac{R_k(\tau)}{S(\tau)^3}\left[T_{ij}\bigl(\bhR(\tau)\bigr)-\delta_{ij}\right]
-\frac{2 c^2\tau^2-R(\tau)^2}{R(\tau)^3 S(\tau)}T_{ijk}\bigl(\bhR(\tau)\bigr)\right\},
\end{align}
\end{subequations}
in which the wave speed $c$ is generic. Let us introduce the notation
\begin{align}
J^{(p)}_{ijk}(\br,t)=J_{ijk}(\br,t)_{c\to c_p}
\end{align}
to denote the tensor $J_{ijk}(\br,\tau)$ in which $c$ is substituted by the specific wave speed $c_p$ with $p={\rm T}$, or $p={\rm L}$. Expressed in terms of $J^{(p)}_{ijk}(\br,t)$, Eq.\ (\ref{eq:iisoijkdef}) reads
\begin{align}
\label{eq:Iisoijk}
&I^{\text{iso}}_{ijk}(\br+\bV t,t)=I^{\text{iso}}_{zzk}(\br+\bV t,t)\delta_{ij}\nonumber\\
&+\frac{1}{2\pi\rho}\Re\sum_{p={\rm T},{\rm L}}\frac{(\pm)}{c_p^2}\left[J^{(p)}_{ijk}(\br-\ii\varepsilon\bbeta^{(p)},t+\ii\varepsilon/c_p)
-J^{(p)}_{ijk}(\br-\ii\varepsilon\bbeta^{(p)},0^++\ii\varepsilon/c_p)\right].
\end{align}

Directly carrying out integral (\ref{eq:jijkgen2}) to obtain $J_{ijk}$ is of considerable difficulty. In view of Eq.~(\ref{eq:jijkgen1}), we instead compute it as the second symmetrized gradient of the following simpler indefinite integral that we verify in
\ref{sec:verifjint}:
\begin{subequations}
\label{eq:prim2}
\begin{align}
\label{eq:prim2e10}
J_j(\br,\tau)&=c\int \dd \tau\, S(\tau)\frac{R_j(\tau)}{R(\tau)^2}\\
\label{eq:prim2e1}
&=\frac{1}{\beta^2}
\left\{S(\tau)\beta_j+\gamma\log[\gamma D(\tau)+S(\tau)] r_m \Aminus_{mj}(\bbeta)-r_m L_l(\tau) B_{mlj}(\bhn)\right\}+\text{TIT},
\end{align}
where $\Aminus_{ij}$ is defined in Eq.\ (\ref{eq:apm}), $B_{mlj}(\bhn)$ is the third-rank tensor
\begin{align}
\label{eq:prim2e2}
B_{mlj}(\bhn)=\bigl(2\hn_m\hn_l-\delta_{ml}\bigr)\delta_{j1}
+\bigl(\epsilon_{zmp}\hn_p\hn_l+\epsilon_{zlp}\hn_p\hn_m\bigr)\delta_{j2},
\end{align}
and $L_i(\tau)$ is the vector
\begin{align}
\label{eq:Ldef}
\mathbf{L}(\tau)=\left(
\begin{array}{c}
{\displaystyle\log\frac{c\tau-S(\tau)}{R(\tau)}}\\
{\displaystyle -\arctan\frac{\br\cdot\bR(\tau)}{(\br\times\bbeta)_z S(\tau)}}
\end{array}
\right).
\end{align}
\end{subequations}
In the latter expressions, $\epsilon_{ijk}$ is the Levi-Civita symbol, and $(\br\times\bbeta)_z=\epsilon_{zkl}r_k\beta_l$. Equation (\ref{eq:prim2e1}) holds up to an arbitrary time-independent and $\br$-dependent integration `constant' in the right-hand side, which explains why the latter blows up as $\beta\to 0$ (i.e., $\bV\to \mathbf{0}$). We call such arbitrary quantities time-independent terms (TITs). Contributions from TITs are ignorable, since they ultimately cancel out in the difference of boundary values in (\ref{eq:Iisoijk}). For the sake of completeness, we provide in \ref{sec:finitelimit} another form of $J_j(\br,t)$ (not used hereafter), in which the integration constant is adjusted to yield the expected finite limit as $\bV\to \mathbf{0}$.

The contraction $L_l B_{ilj}$ in (\ref{eq:prim2e1}) can be made more transparent by computing the components of $B_{ilj}$. One finds that $B_{1l1}(\bhn)=-T_{l1}(\bhn)$, $B_{2l2}(\bhn)=T_{l1}(\bhn)$ and $B_{1l2}(\bhn)=B_{2l1}(\bhn)=-T_{l2}(\bhn)$. Thus, introducing the following traceless matrices that correspond to pure-shear and simple-shear modes of plane-strain deformation (e.g., \cite{WILL08}):
\begin{align}
\mathsf{U}^{(1)}=\left(\begin{array}{cc}1&0\\0&-1\end{array}\right)\qquad
\mathsf{U}^{(2)}=\left(\begin{array}{cc}0&1\\1&0\end{array}\right),
\end{align}
one has
\begin{subequations}
\label{eq:bidentities}
\begin{align}
B_{i1j}(\bhn)&=-\sum_{\alpha=1,2}T_{1\alpha}(\bhn)U^{(\alpha)}_{ij}=-T_{ij}(\bhn),\\
B_{i2j}(\bhn)&=-\sum_{\alpha=1,2}T_{2\alpha}(\bhn)U^{(\alpha)}_{ij}=-\epsilon_{zip}T_{pj}(\bhn),
\end{align}
\end{subequations}
and
\begin{align}
L_l B_{ilj}=-\sum_{\alpha=1,2}L_l T_{l\alpha}(\bhn)U^{(\alpha)}_{ij}.
\end{align}

We can now obtain $J_{ijk}$ as the second symmetrized gradient of $J_j(\br,\tau)$. In view of definition (\ref{eq:jijkgen1}) of  $J_{ijk}$ by an integral, this way of proceeding involves interchanging the integral and the
space derivatives in this definition. Strictly speaking however,
\begin{align}
\label{eq:jsing}
\frac{1}{2}\left[\partial^2_{ik}J_{j}(\br,\tau)+\partial^2_{jk}J_{i}(\br,\tau)\right]=J_{ijk}(\br,\tau)+\text{Sing.},
\end{align}
where we emphasize the fact that the left-hand side is equal to $J_{ijk}(\br,\tau)$ only up to time-\emph{dependent} singular Dirac-like distributional parts concentrated on the source path. Indeed, Eq.~(\ref{eq:tddirac}) in
\ref{sec:darctan} shows that such a Dirac term, of support the line $\br^\perp=\mathbf{0}$, arises when differentiating the arctangent in (\ref{eq:Ldef}), because the latter function is a discontinuous function of $\br$. This can be traced to the fact that the integrand in definition (\ref{eq:prim2e1}) of $J_j$ is inversely proportional to the norm of $\bR(\tau)=\br+\bV\tau$, which vanishes  at time $\tau=-\br\cdot\bhV/V$ when $\br^\perp=\mathbf{0}$. Since $\bR$ is invariant under the complex-valued shifts considered in our calculations [see remark following (\ref{eq:iisozzk})], this kind of singularity is not removed by analytic continuation. Thus, for instance, the integral
\begin{align}
c\int_{0^++\ii\varepsilon/c}^{t+\ii\varepsilon/c}\dd\tau\,\left[S(\tau)\frac{R_j(\tau)}{R(\tau)^2}
\right]_{\br\to\br-\ii\varepsilon\bbeta},
\end{align}
whose second derivative enters (\ref{eq:iisoijkdef}), is ill-defined (infinite) on the segment $\mathcal{L}(t)=\{\br\,|\, \br=-\tau\bV, \tau\in(0,t)\}$ contained in the glide plane $\br^\perp=0$, which represents the trajectory of the source expressed in the co-moving frame. Quite generally, the source path is a notorious source of difficulties (see, e.g., \cite{GURR13} and references therein).

Here however, the difficulty is somewhat artificial, as it stems from the route adopted to reach $J_{ijk}(\br,\tau)$. For brevity, we bypass the problem by remarking that the latter is necessarily a function without any Dirac term, since by (\ref{eq:jijkgen2}) its time-derivative is the integrand, which is free of Dirac-like singularities. It follows that extracting $J_{ijk}(\br,\tau)$ from a double differentiation of $J_j(\br,\tau)$ can be done simply by \emph{dropping any Dirac-like distributional parts that would normally arise in the process of differentiating the arctangent} in Eq.\ (\ref{eq:Ldef}).

Then, completing the calculation reduces to carrying out cumbersome differentiations. Introducing new vectors
\begin{align}
\bQ=\bR/S,\qquad \bq=\br/(\br\cdot\sfAplus\cdot\br),
\end{align}
one has
\begin{align}
\partial_i S=-Q_i,\qquad \partial_i \gamma\log(\gamma D+S)=(q_i\beta_l-\beta_i q_l)Q_l-\frac{c\tau}{S}q_i+\text{TIT},
\end{align}
where an ignorable term in the second expression has been lumped in the TIT notation. In terms of these gradients, the first symmetrized derivative of $J_i$ reads
\begin{align}
\frac{1}{2}(J_{i,j}+J_{j,i})
&=\frac{1}{2\beta^2}
\biggl\{-(\beta_i Q_j+\beta_j Q_i)
+\left[\Aminus_{im}(q_j\beta_l-\beta_j q_l)+\Aminus_{jm}(q_i\beta_l-\beta_i q_l)\right]r_m Q_l\nonumber\\
&-\frac{c\tau}{S}\left[\Aminus_{im}q_j+\Aminus_{jm}q_i\right]r_m\nonumber\\
&+2\gamma\log(\gamma D+S)\Aminus_{ij}-2 L_m B_{imj}-r_m\left(L_{l,i}B_{mlj}
+L_{l,j}B_{mli}\right)\biggr\}+\text{TIT},
\end{align}
where gradients of the components of vector $\mathbf{L}$ defined by (\ref{eq:Ldef}) are as follows:
\begin{subequations}
\begin{align}
\partial_i L_x&=\frac{c\tau}{S}\frac{R_i}{R^2},\\
\label{eq:dly}
\partial_i L_y&=-\epsilon_{zip}\left[\frac{c\tau}{S}\frac{R_p}{R^2}+\left(\bbeta\cdot\bQ-\frac{c\tau}{S}\right)q_p\right].
\end{align}
\end{subequations}
In arriving at (\ref{eq:dly}) the following identity has been employed:
\begin{align}
\left[(\br\times\bbeta)_z S\right]^2+(\br\cdot\bR)^2=(\br\cdot\sfAplus\cdot\br)\,R^2.
\end{align}
Unless the TIT is suitably chosen in Eq.~(\ref{eq:prim2e1}) (which we shall not bother to do), $J_{j,i}$ is not a symmetric tensor. This lack of symmetry is our motivation for having symmetrized Eq.\ (\ref{eq:gijdgjeq1}) in the first place.

Introducing further
\begin{align}
U_{ij}=\delta_{ij}+Q_i Q_j,\qquad V_{ij}=\frac{\Aminus_{ij}}{\br\cdot\Aplus\cdot\br},\qquad W_{ij}=\delta_{ij}-2\frac{r_m \Aplus_{mi}r_j}{\br\cdot\Aplus\cdot\br},
\end{align}
in terms of which
\begin{align}
\partial_k Q_i=\frac{1}{S}U_{ki},\qquad \partial_k q_i=\frac{W_{ki}}{\br\cdot\sfAplus\cdot\br},
\end{align}
the second symmetrized derivative (\ref{eq:jsing}), amputated from its singular part, finally provides
\begin{align}
J_{ijk}(\br,\tau)&=
\frac{1}{2\beta^2}
\biggl\{-\frac{1}{S}(U_{kj}\beta_i+U_{ki}\beta_j)\nonumber\\
&+\left[(W_{ki}\beta_l-\beta_i W_{kl})V_{mj}+(W_{kj}\beta_l-\beta_j W_{kl})V_{mi}\right]r_m Q_l\nonumber\\
&{}+\left[(r_i\beta_l-\beta_i r_l)V_{kj}+(r_j\beta_l-\beta_j r_l)V_{ki}+2(r_k \beta_l-\beta_k r_l)V_{ij}\right]Q_l\nonumber\\
&{}+\frac{1}{S}\left[V_{im}(r_j\beta_l-\beta_j r_l)+V_{jm}(r_i\beta_l-\beta_i r_l)\right]r_m U_{kl}\nonumber\\
&{}-\frac{c\tau}{S}\biggl[\frac{Q_k}{S}\left(V_{im}r_j+V_{jm}r_i\right)r_m+\left(V_{ik}r_j+V_{jk}r_i+2 V_{ij}r_k\right)\nonumber\\
&+\left(V_{im}W_{kj}+V_{jm}W_{ki}\right)r_m\biggr]\nonumber\\
\label{eq:symder2}
&{}-(L_{l,i}B_{klj}+L_{l,j}B_{kli}+2 L_{l,k} B_{ilj})
-r_m\left(L_{l,ik}B_{mlj}+L_{l,jk}B_{mli}\right)
\biggr\}+\text{TIT},
\end{align}
where
\begin{subequations}
\label{eq:Lxy}
\begin{align}
L_{x,ik}&=\frac{c\tau}{S R^2}\left(U_{ik}-2\hR_i\hR_k\right),\\
L_{y,ik}&=-\epsilon_{zip}\biggl[L_{x,pk}+\frac{1}{S}\left(U_{kl}\beta_l-\frac{c\tau}{S}Q_k\right)q_p
+\left(\bbeta\cdot\bQ-\frac{c\tau}{S}\right)\frac{W_{kp}}{\br\cdot\sfAplus\cdot\br}\biggr]
+\text{TIT}.
\end{align}
\end{subequations}
The zero-velocity form of $J_{ijk}(\br,\tau)$ is most easily obtained, not from Eq.\ (\ref{eq:symder2}), but rather by differentiating twice expression (\ref{eq:Jzerovel}) of $J_j(\br,t)|_{\bV=\mathbf{0}}$. One finds
\begin{align}
J_{ijk}(\br,\tau)|_{\bV=\mathbf{0}}=\frac{c\tau}{r\sqrt{c^2\tau^2-r^2}}\,\hr_i\hr_j\hr_k
-\frac{c\tau}{r^3}\sqrt{c^2\tau^2-r^2}\,T_{ijk}(\bhr)+\text{TIT}.
\end{align}
This expression blows up when $\tau\to\infty$, but this is not a problem. Indeed the part that explodes, namely $-(c^2\tau^2/r^3)T_{ijk}(\bhr)$, cancels out in expressions such as (\ref{eq:Iisoijk}) where it is eliminated [after division by the relevant wave speed squared; see (\ref{eq:Iisoijk})] from adding the longitudinal and transverse contributions, which have opposite signs.

Substituting $J_{ijk}(\br,\tau)$ and $I^{\text{iso}}_{ijk}(\br+\bV t,t)$ as given by (\ref{eq:Iisozzk}) into Eq.\ (\ref{eq:Iisoijk}) completes our solution for $I^{\text{iso}}_{ijk}(\br+\bV t,t)$. The quantity $I^{\text{iso}}_{ijk}(\br,t)$ follows from going back to the reference frame by means of the substitution $\br\to\br -\bV t$. The analysis of the distributional limit $\varepsilon\to 0$ is too long to be reported, but would follow the same lines as in the antiplane-strain case.

\begin{figure}[!ht]
\centering
\includegraphics[width=15.1cm]{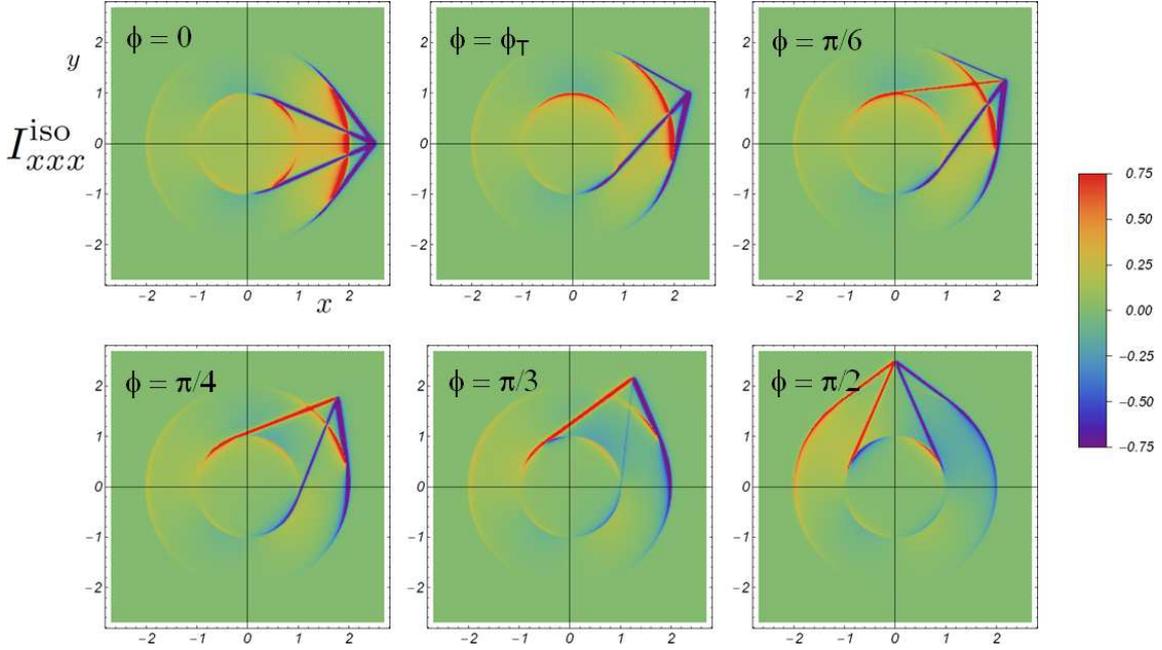}
\caption{\label{fig:fig6} Plane strain problem. Component $I^{\text{iso}}_{xxx}(\br,t)$ for supersonic velocity $V=2.5\,\cT$, and various orientations of the velocity vector (as indicated). Units are such that $\cT=1$ and $\mu=1$, $\cL=2\cT$, and $\varepsilon=0.01$.}
\end{figure}
Fig.\ \ref{fig:fig6} illustrates these plane-strain calculations on the component $I^{\text{iso}}_{xxx}(\br,t)$, for supersonic motion at velocity $|V|>\cL$, with focus on the branch-extinction property. The outer circular wave is of longitudinal character; the inner one is transverse. Various branches of either Mach cone (or both) can be extinguished, or almost extinguished, on any component of the tensor $I_{ijk}$ for specific orientations of the velocity. In the figure the fundamental critical angles for Mach-cone branch extinction are $\phi_{\rm T}=\arctan(\betaT^2-1)^{-1/2}\simeq 0.411$ (shear cone) and $\phi_{\rm L}=\arctan(\betaL^2-1)^{-1/2}\simeq 0.923$ (longitudinal cone) radians. For an angle such as $\phi=\pi/3$ close ---but not \emph{very} close--- to these critical angles or to one of their symmetric counterparts (see previous section), a simultaneous quasi-extinction of branches on both cones is observed.

\section{Summary and discussion}
\label{sec:concl}
To summarize, we emphasized the often overlooked distributional character of the Green tensors of the plane-strain and anti-plane strain components of the Navier equation, and their gradients, and studied a regularization procedure for these quantities. The regularized expressions are suitable to direct exploitation in practical problems involving nonuniformly moving source lines. The regularization consists in spreading the point source over a finite width $\varepsilon$, by means of an analytic continuation of the elastodynamic Green tensor to complex time. The connection between this procedure and one of convolution by one particular source shape function has been established.

As an application, some important definite integrals of tensor fields over finite time intervals were obtained from  non-trivial \emph{indefinite} integrals, for both plane-strain and anti-plane-strain configurations. These definite integrals, to be used elsewhere \cite{LAZA14}, are encountered in the problem of a moving line source suddenly accelerated from rest to a constant velocity. In this context, they determine the material velocity and strain fields.
Numerical computations were carried out to illustrate the fact that, since the Green function embodies all wave propagation effects, cases of subsonic, transonic, or supersonic source motion can be handled altogether by one single closed-form expression. Moreover, a property (of geometric origin) of Mach-cone branch extinction on Cartesian components of the tensor integrals under consideration was reported.

There are two ways of envisioning the obtained integrals. The first one is to consider them as computational intermediates to reach the fields in the form of distributions. Indeed, the method is straightforward since we do not need to care about causality constraints at wavefronts. As was explicitly demonstrated in the anti-plane case, the distributional forms are obtained by letting the source width parameter $\varepsilon\to 0$ (Volterra limit). In this limit, the resulting distributions represent fields generated by the moving point source. However (see Introduction), such distributional Volterra solutions are devoid of any physical meaning in the vicinity of their singularities, namely, the source position and the wavefronts. This is particularly obvious for faster-than-wave motions where Mach cones show up in the limit in terms of the Dirac distribution. Thus, no meaningful numbers can be extracted from the limiting integrals, unless the latter are convoluted with a suitable source shape function to produce physical fields. Of course, using for this purpose the isotropic shape function employed for regularizing the Green tensor from the outset would result in the same expressions with $\varepsilon$ finite reported above (assuming the calculation to be manageable this way, which is far from granted; this is why we started from Green's function instead). Other shape functions could in principle be employed, perhaps at the price of additional difficulties.

The alternative point of view ---which has our preference--- is to consider that the isotropic shape function we employed \emph{is} of physical relevance (at least approximately, say, as some effective mean core shape; e.g., in the sense of an angular averaging), and to use for simulations the obtained definite integrals as they stand. This spreads Mach cones over a width of order $\varepsilon$, and makes fields finite everywhere. Keeping $\varepsilon$ nonzero reveals that dramatic variations in strength \emph{and sign} of physical fields take place near wavefronts, as our figures clearly show, which may be of consequence for the dislocation nucleation problem \cite{GURR13}.

Then, there remains the question of the value to be attributed to $\varepsilon$. From a purely numerical standpoint, any value not leading to overflows in field expressions is acceptable. On the other hand, physics requires \emph{both} the source position \emph{and} its width to be prescribed by coupled equations of motion, which makes $\varepsilon(t)$ a function of time \cite{PELL13}. For dislocations in metals, this ``physical'' $\varepsilon$ does not exceed a few multiples of the interatomic distance (but can also be notably smaller, depending on the velocity and acceleration regimes). From this perspective, the definite integrals obtained above must be though of as a post-processing device to estimate fields in the surrounding medium, once the dislocation position and ``physical" width are known as functions of time; the typical size $\varepsilon$ considered in the present work being substituted by the ``physical" one.\footnote{Ideally, the regularizing source shape should, for consistency, be the same as that employed in deriving the equations of motion of the source. This would be asking for too much in the present state of knowledge (the equations of motion of Ref.\ \cite{PELL13} were obtained within harmonic regularization).} However, it is important to point out that our present derivations do not account for the specific radiative contributions induced by such a time-dependence of $\varepsilon$. With applications in mind, it could nonetheless be feasible to tentatively neglect the latter (assuming it to be subdominant) and carry out this substitution within each time interval in a numerical procedure of the type alluded to in the Introduction.

\section*{Acknowledgments}
M.L.\ gratefully acknowledges the grants obtained from the Deutsche Forschungsgemeinschaft (Grant Nos.\ La1974/2-1, La1974/2-2, La1974/3-1). The authors thank one anonymous reviewer for useful remarks.

\appendix
\section{Differentiation in the complex plane in the sense of distributions}
\label{sec:dcp}
We argue here that differentiation in the sense of distributions of principal determinations of logarithms or square roots in the complex plane must be performed bearing in mind that the derivative involves a distributional part. The latter is supported by the negative real half-axis, on which the principal determinations undergo a jump in the direction parallel to the imaginary axis. Thus,
\begin{subequations}
\begin{align}
\label{eq:dlogz}
\dd \log z
&=\frac{\dd z}{z}+2\ii\pi \theta(-\Re z)\delta(\Im z)\dd (\Im z),\\
\dd \sqrt{z}&=\frac{\dd z}{2\sqrt{z}}+2\ii (-\Re z)^{1/2}_+\delta(\Im z)\dd (\Im z),\\
\label{eq:zmonehalf}
\dd\, z^{-1/2}&=-\frac{1}{2}z^{-3/2}\dd z-2\ii (-\Re z)^{-1/2}_+\delta(\Im z)\dd (\Im z).
\end{align}
\end{subequations}
As a rule, differentiation, with respect to some variable $u$, of the logarithm or square root of a complex-valued function $z=f(u)$ creates a Dirac-like distributional part, if: (i) the imaginary part of $f(u)$ depends on $u$, and (ii) there exists values of $u$ such that $\Re f(u)<0$ while $\Im f(u)=0$. For instance, from Eq.\ (\ref{eq:zmonehalf}) follows the non-trivial result that
\begin{align}
\frac{\partial}{\partial t}\frac{1}{\sqrt{(c t+\ii\varepsilon)^2-r^2}}
=-\frac{c(c t+\ii\varepsilon)}{[(c t+\ii\varepsilon)^2-r^2]^{3/2}}-\frac{2\ii}{\sqrt{\varepsilon^2+r^2}}\delta(t)
\end{align}
(this equation is not needed in the main text).

\section{Proof of Equation (\ref{eq:prim2e1})}
\label{sec:verifjint}
The proof proceeds by differentiation. One finds, in succession
\begin{subequations}
\begin{align}
&\frac{1}{c}\frac{\dd}{\dd\tau}S(\tau)=\frac{D}{S},\\
&\frac{1}{c}\frac{\dd}{\dd\tau}D(\tau)=1-\beta^2,\\
&\frac{1}{c}\frac{\dd}{\dd\tau}\log[\gamma D(\tau)+S(\tau)]=1/(\gamma S),\\
&\frac{1}{c}\frac{\dd}{\dd\tau}\log\frac{c\tau-S(\tau)}{R(\tau)}=-\frac{1}{S}\frac{\br\cdot\bR}{R^2},\\
\label{eq:dtarctan}
&-\frac{1}{c}\frac{\dd}{\dd\tau}\arctan\frac{\br\cdot\bR(\tau)}{(\br\times\bbeta)_z S(\tau)}
=\frac{c\tau}{S}\frac{(\br\times\bbeta)_z}{R^2}.
\end{align}
\end{subequations}
It follows that
\begin{align}
\frac{\dd}{\dd\tau} J_k(\br,\tau)
&=\frac{D\beta_k}{\beta^2 S}
+\frac{r_i \Aminus_{ik}(\bbeta)}{\beta^2 S}
-\frac{r_i}{\beta^2 S R^2}
\left[-\br\cdot\bR\,\delta_{j1}+c\tau(\br\times\bbeta)_z\delta_{j2}\right]
B_{ijk}(\bhn)\nonumber\\
&=\frac{D\beta_k}{\beta^2 S}
+\frac{r_i \Aminus_{ik}(\bbeta)}{\beta^2 S}
-\frac{r_i}{\beta^2 S R^2}
\left[(\br\cdot\bR)T_{ik}(\bhn)-c\tau\,\epsilon_{zmn} r_m \beta_n\epsilon_{zip}T_{pk}(\bhn)\right],
\end{align}
where the last form results from Eqs.\ (\ref{eq:bidentities}). Moreover, since indices $m,n,i,p$ take on values $1,2$,
\begin{align}
\epsilon_{zmn}\epsilon_{zip}=\delta_{im}\delta_{np}-\delta_{mp}\delta_{in}.
\end{align}
Therefore,
\begin{align}
\beta^2 S R^2\frac{\dd}{\dd\tau} J_k(\br,\tau)
&=R^2\left[D \beta_k+r_i\Aminus_{ik}(\bbeta)\right]
-r_i\left[(\br\cdot\bR)T_{ik}(\bhn)-ct(r_i\beta_p-r_p\beta_i)T_{pk}(\bhn)\right].
\end{align}
A straightforward expansion of the right-hand side then shows that
\begin{align}
\beta^2 S R^2\frac{\dd}{\dd\tau} J_k(\br,\tau)=\beta^2 S^2 R_k,
\end{align}
whence the result.

\section{Alternative indefinite integral for the plane-strain case}
\label{sec:finitelimit}
By going to the limit $\bV\to\mathbf{0}$ in the integrand, the integral in (\ref{eq:prim2e10}) naturally reduces to
\begin{align}
\label{eq:Jzerovel}
J_j(\br,\tau)|_{\bV=\mathbf{0}}&=c\int \dd \tau\, \sqrt{c^2 \tau^2-r^2}\frac{\br}{r^2}\nonumber\\
&=\frac{\br}{2r^2}\left[c \tau\sqrt{c^2\tau^2-r^2}-r^2\log\left(c\tau+\sqrt{c^2\tau^2-r^2}\right)\right]+\text{Const.}
\end{align}
The arbitrary implicit time-independent constant in (\ref{eq:prim2e1}) can be adjusted to retrieve this limit, by employing the same equation, but now with
\begin{align}
\mathbf{L}=\left(
\begin{array}{c}
{\displaystyle\log\frac{c\tau-S}{R}-\log r-\frac{1}{2}(\bhr\cdot\bbeta)^2}\\
{\displaystyle
-\arctan\frac{\br\cdot\bR}{(\br\times\bbeta)_z S}+\frac{\pi}{2}\sign(\br\times\bbeta)_z
}
\end{array}
\right).
\end{align}
It should be noted that as the regularization considered in the paper involves adding to $\br$ an imaginary part proportional to $\bbeta$ [see Eq.\ (\ref{eq:Iisoijk})], the quantity $(\br\times\bbeta)_z$ is always a real number. Although it is more regular (the expression of $L_y$ being now continuous as $\bbeta\to 0$), the above expression of $\mathbf{L}$ differs from (\ref{eq:Ldef}) by ignorable TITs.

\section{Distributional derivative of the arctangent in $\mathbf{L}$}
\label{sec:darctan}
We clarify hereafter the issue of the distributional derivative of the arctangent in $\mathbf{L}$. First, it is recalled that $\arctan x$ is a continuous function, while $\arctan(1/x)$ is discontinuous at $x=0$. The discontinuity manifests itself in the following well-known identity:
\begin{align}
\arctan x+\arctan \frac{1}{x}=\frac{\pi}{2}\sign x.
\end{align}
It follows that, in the distributional sense,
\begin{align}
\left(\arctan\frac{1}{x}\right)'=\left(\frac{\pi}{2}\sign x-\arctan x\right)'=\pi\delta(x)-\frac{1}{1+x^2}.
\end{align}
This distributional derivative differs from the usual one by the presence of the Dirac term in the right-hand side, and plays an important role when working with single-valued material displacements generated by dislocations \cite{PELL10}.

Consider now $h(x)=\arctan[f(x)/g(x)]$ where both functions $f$ and $g$ change signs at their zeros. In view of the above derivative, a little reflection shows that the derivative $h'(x)$ must be computed in the sense of distributions as
\begin{align}
h'(x)&=\left.\frac{\dd}{\dd x}\arctan\frac{f(x)}{g(x)}\right|_g
+\left.\frac{\dd}{\dd x}\left\{\frac{\pi}{2}\sign[f(x)]\sign[g(x)]-\arctan\frac{g(x)}{f(x)}\right\}\right|_f\nonumber\\
\label{eq:dxarctan}
&=\frac{f'(x)g(x)-g'(x)f(x)}{f^2(x)+g^2(x)}+\pi\sign[f(x)]g'(x)\delta\big(g(x)\big),
\end{align}
where the subscripts indicate that differentiations are carried out with $g$ and $f$ kept constant, respectively. The Dirac peaks in the derivative at the zeros of $g(x)$ are observable numerically when $h'(x)$ is approximated by
$[h(x+\eta)-h(x-\eta)]/(2\eta)$, with $\eta$ a small number.

Applying this formula to the differentiation of $L_y$ in (\ref{eq:Ldef}) yields
\begin{align}
\partial_i L_y&=-\partial_i\arctan\frac{\br\cdot\bR}{(\br\times\bbeta)_z S}\nonumber\\
&=-\frac{S}{R^2}
\left\{(\bq\times\bbeta)_z(2 S+\br\cdot\bQ) Q_i-\left[c\tau (\bq\times\bbeta)_z\beta_i+\bq\cdot\bR \epsilon_{zim}\beta_m\right]\right\}\nonumber\\
\label{eq:tddirac}
&-\pi\sign(\br\cdot\bR)(\epsilon_{zim}\beta_m)\delta\bigr((\br\times\bbeta)_z\bigl).
\end{align}
Up to TITs, the regular part can be reduced to (\ref{eq:dly}), while the Dirac part, supported by the line $\br^\perp=0$, is not time-independent since it involves $\bR=\bR(\tau)$ in its prefactor. This is the source of the
difficulty encountered in Section \ref{sec:applpsp}.

\end{document}